\documentclass[fleqn,10pt]{wlscirep}
\usepackage[utf8]{inputenc}
\usepackage[T1]{fontenc}
\usepackage{verbatim}

\title{Scattering polarimetry \textcolor{black}{enables} correlative nerve fiber imaging and multimodal analysis}

\author[1]{Franca auf der Heiden}
\author[1,2,*]{Markus Axer}
\author[1,3]{Katrin Amunts}
\author[4,1,*]{Miriam Menzel}
\affil[1]{Institute of Neuroscience and Medicine (INM-1), Forschungszentrum Jülich GmbH, 52425 Jülich, Germany}
\affil[2]{Department of Physics, School of Mathematics and Natural Sciences, University of Wuppertal, 42119 Wuppertal, Germany}
\affil[3]{C. and O. Vogt Institute for Brain Research, University Hospital Düsseldorf, 40225 Düsseldorf, Germany}
\affil[4]{Department of Imaging Physics, Delft University of Technology, 2628 CJ Delft, The Netherlands}
\affil[*]{m.axer@fz-juelich.de, m.menzel@tudelft.nl}


\begin{abstract}
	Mapping the intricate network of nerve fibers is crucial for understanding brain function. Three-Dimensional Polarized Light Imaging (3D-PLI) and Computational Scattered Light Imaging (ComSLI) map dense nerve fibers in brain sections with micrometer resolution using visible light. 3D-PLI reconstructs 3D-fiber orientations, while ComSLI disentangles multiple directions per pixel. So far, these imaging techniques have been realized in separate setups. A combination within a single device would facilitate faster measurements, pixelwise mapping, cross-validation of fiber orientations, and leverage the advantages of each technique while mitigating their limitations. Here, we introduce the Scattering Polarimeter, a \textcolor{black}{microscope} that facilitates correlative large-area scans by integrating 3D-PLI and ComSLI measurements into a single system. Based on a \textcolor{black}{Mueller} polarimeter, it incorporates variable retarders and a large-area light source for direct and oblique illumination, enabling \textcolor{black}{combined} 3D-PLI and ComSLI measurements\textcolor{black}{.} Applied to human and vervet monkey brain sections, the Scattering Polarimeter generates results comparable to state-of-the-art 3D-PLI and ComSLI setups and creates a multimodal fiber direction map, integrating the robust fiber orientations obtained from 3D-PLI with fiber crossings from ComSLI. Furthermore, we discuss applications of the Scattering Polarimeter for unprecedented correlative and multimodal brain imaging.
\end{abstract}
\begin{document}

\flushbottom
\maketitle
%
%
\thispagestyle{empty}


\section*{Introduction}
The human brain is a highly complex system, consisting of about 86$\times10^9$ neurons and about as many glial cells~\cite{Bartheld2016}. Together, they form an intricate system in which a neuron can make up to $10^5$ synaptic connections with other neurons~\cite{Snell2010}. A major task in neuroscience is the investigation of this entangled network. Nowadays, the spectrum of neuroimaging techniques for mapping nerve fiber architectures spans several orders of magnitude in resolution. However, achieving high resolution often comes at the expense of a comprehensive spatial overview due to practical limitations in measurement and computation time.

Diffusion-weighted magnetic resonance imaging (dMRI or DWI) allows the in-vivo study of a whole brain with a resolution of about one millimeter and is mainly limited by scanning time and motion artifacts~\cite{Yendiki2022}. Post-mortem studies have achieved a resolution down to 150\,µm~\cite{Calabrese2015}. The probabilistic distribution of axonal orientations can be estimated from dMRI data~\cite{Shi2017}. However, the measurement of crossing fibers poses significant challenges, often resulting in false positives during tractography~\cite{MaierHein2017}.
On the other side of the scale, fluorescence microscopy (FM), and transmission electron microscopy (TEM) offer resolutions down to several nanometers, thereby enabling the subcellular investigation of neurofilaments and axonic myelin sheaths. When combined with super-resolution microscopy approaches, FM circumvents the diffraction limit of light microscopy~\cite{Sanderson2014}. TEM can resolve nanometer-scale structures from tissue stained with heavy metals (typically osmium), e.g.\,the detailed structure of neurons or the lipid bilayers of the myelin~\cite{Franken2020}. Due to practical constraints, super-resolution microscopy is restricted to small sample areas and cannot provide an overview of larger structures, such as long fiber tracts.

In-between, light microscopy achieves a resolution down to the diffraction limit of about one micrometer while providing an overview over whole brain sections. In modern neuroscience, light microscopy benefits from advanced tissue preparation and staining methods, automation for faster image acquisition, and increased computational power~\cite{Osten2013}.
However, standard microscopy techniques cannot trace the course of fiber pathways in dense tissues as they yield similar contrast for nerve fibers with different orientations. To reconstruct the dense fiber network of the brain, techniques that directly visualize fiber orientations need to be employed: Polarimetric techniques like \textcolor{black}{Mueller} polarimetry~\cite{Novikova2022, Felger2023} or Three-Dimensional Polarized Light Imaging (3D-PLI)~\cite{Axer2011, Axer2011a} exploit the anisotropic refraction (birefringence) of the nerve myelin sheaths to determine the fiber orientations. Computational Scattered Light Imaging (ComSLI)~\cite{Menzel2021, Menzel2021a}, on the other hand, measures the anisotropic scattering of light on directed structures such as nerve fibers. In contrast to standard light microscopy, these techniques do not require any staining and can be applied to study the course of long-range fiber pathways in unstained histological brain sections.

The techniques have different advantages and disadvantages: 
\textcolor{black}{Mueller polarimetry} uses different combinations of polarization states to derive a \textcolor{black}{Mueller matrix} for each image pixel by solving a linear equation system. Decomposition techniques, such as the Lu-Chipman decomposition~\cite{Lu1996} \textcolor{black}{(see Supplementary 1), rewrite the Mueller matrix as product of depolarizer, retarder, and diattenuator}, allowing the measurement of all optical properties of a sample. Additionally, \textcolor{black}{Mueller} polarimetry can distinguish between linear and circular optical effects.
\textcolor{black}{3D-PLI can be considered as incomplete Mueller polarimetry and be used to study (linear) retardance of tissue samples. It is faster than full Mueller polarimetry and yields more stable signals, but does not provide any information about depolarization,  diattenuation, or circular optical effects.}
3D-PLI measures the birefringence of a brain section \textcolor{black}{and} provides information about the in-plane direction of the nerve fibers, and about the out-of-plane inclination which is related to the strength of the birefringence signal (retardation). 
\textcolor{black}{3D-PLI provides robust results for fiber orientations, also for regions with lower fiber density like gray matter in the cortex.}
However, \textcolor{black}{both Mueller polarimetry and 3D-PLI} cannot resolve fiber crossings, as \textcolor{black}{they only detect} the average signal across multiple crossing directions. \textcolor{black}{Supplementary 1 and 2 provide a more detailed description of how linear retardance and fiber direction are computed using Lu-Chipman decomposition and 3D-PLI signal analysis, respectively.}
ComSLI is based on light scattering at nerve fibers: The sample is illuminated from multiple angles, and the vertically scattered/transmitted light is measured. This technique provides information about fiber directions and inclinations\textcolor{black}{, and} can resolve fiber crossings, determining multiple fiber orientation per image pixel. \textcolor{black}{In contrast to polarization-based techniques, ComSLI does not rely on the briefringence of the sample and can be performed with unpolarized light, allowing its application to differently prepared samples, including formalin-fixed paraffin-embedded brain tissues with different stains \cite{Georgiadis2024}.} However, it is more affected by statistical noise. \textcolor{black}{In principle, ComSLI can provide full scattering patterns for each image pixel \cite{Menzel2021}. A much faster way of measuring fiber directions is angular ComSLI \cite{Menzel2021a} which illuminates the sample under a large polar angle at different azimuthal angles, and derives the fiber orientations from the peak positions in the resulting azimuthal intensity profiles. In the following, the term ComSLI will be used to refer to angular ComSLI.}

\textcolor{black}{As ComSLI relies on a different imaging principle and has different advantages and disadvantages than 3D-PLI or Mueller polarimetry, it complements the other techniques. Combining the scattering and polarimetric results would yield a more reliable and complete reconstruction of the brain's nerve fiber organization. However,}
up to now, the different techniques have been realized in separate setups\textcolor{black}{, making combined measurements and analysis more difficult}. A combination of Mueller polarimetry, 3D-PLI and ComSLI within a single device would facilitate faster measurements, pixel-wise mapping, cross-validation of fiber orientations, and leverage the unique advantages of each technique while mitigating their limitations.

A standard \textcolor{black}{Mueller} polarimeter consists of a polarization state generator (PSG) and a polarization state analyzer (PSA), while a 3D-PLI setup operates as an incomplete \textcolor{black}{Mueller} polarimeter that employs linearly polarized light at different orientations for illumination and a circular analyzer. In contrast, ComSLI employs unpolarized light at large incident angles.
To realize all techniques within a single device, using a shared light source and camera, a specific custom design is needed. Due to the different requirements of the different imaging techniques, it is not possible to simply adapt an existing \textcolor{black}{Mueller} polarimeter.

In this work, we introduce a novel \textcolor{black}{device} -- the \textit{Scattering Polarimeter} -- that facilitates correlative large-area scans by integrating \textcolor{black}{polarimetric and scattering} measurements into a single system \textcolor{black}{(modified Mueller polarimeter)}. To enable direct illumination for the polarimetric measurements and oblique illumination for the scattering measurement, a large LED panel was employed. The PSG and PSA were implemented using liquid crystal variable retarders (LCVRs) and polarization filters, allowing for a fast generation and analysis of all required polarization states without moving elements (cf.\ Fig.\ \ref{fig:pli-results}a). In ComSLI, the PSG elements were bypassed by the oblique illumination (cf.\ Fig.\ \ref{fig:sli-results}a), see Methods for more details. Since polarization effects are negligible in ComSLI, the PSA elements \textcolor{black}{remained} in the optical path without affecting the measurement\textcolor{black}{.}

\textcolor{black}{In principle, the developed Scattering Polarimeter can be used to perform full Mueller polarimetry measurements and study different optical properties of the tissue (see Supplementary 1). However, as the focus of our study is a more reliable reconstruction of nerve fiber orientations, we optimized the device for 3D-PLI, which is a more robust way of deriving fiber orientations than using full Mueller polarimetry.}

In the following, 3D-PLI and ComSLI results are compared to measurements with state-of-the-art setups. Finally, potential applications for multimodal parameter analysis are demonstrated.


\section*{Results}
\begin{figure}[htbp]
	\centering
	\includegraphics[width=\linewidth]{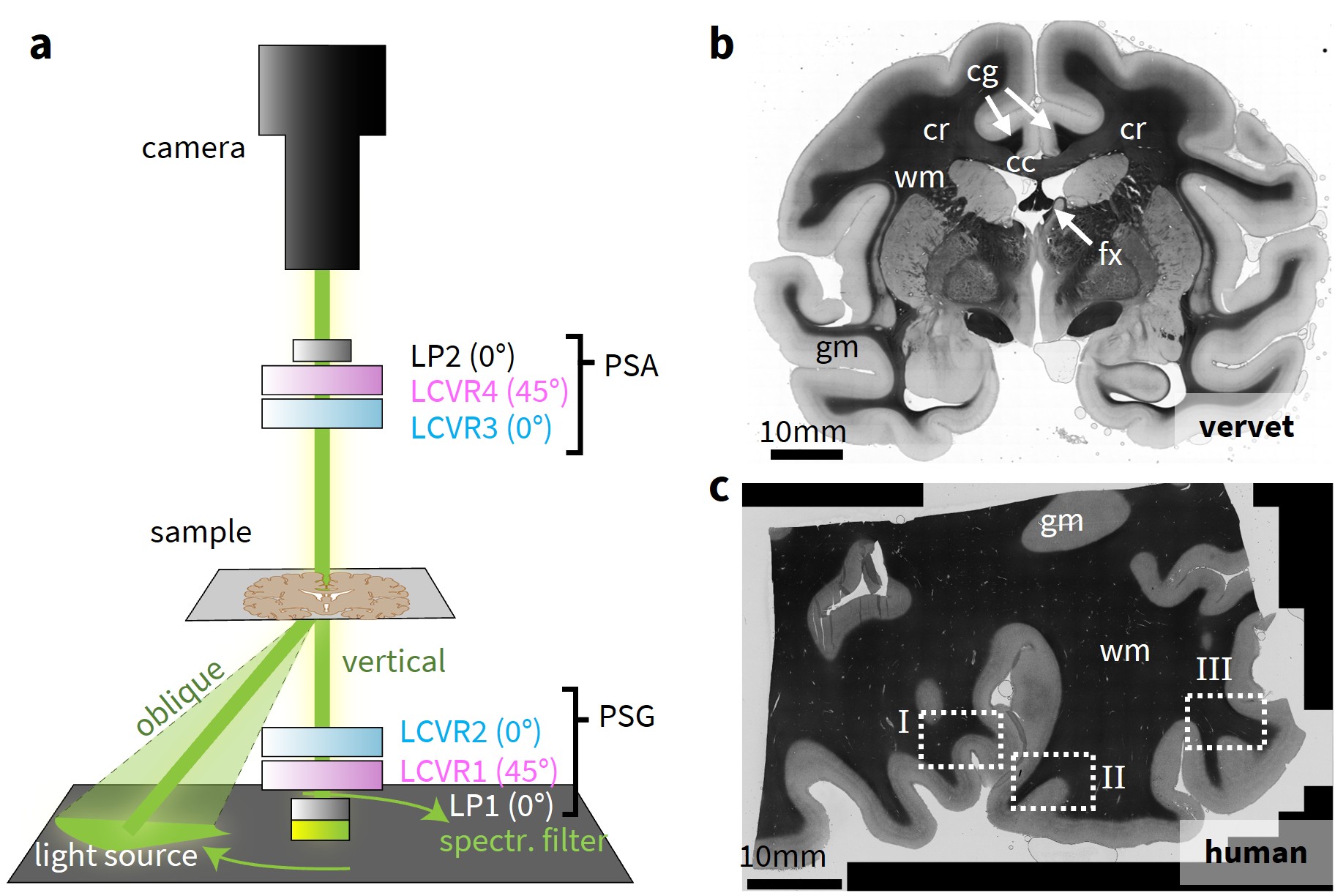}
	\caption{Scattering Polarimeter and brain tissue samples. (\textbf{a}) Setup of the Scattering Polarimeter. The device is realized as a (modified) \textcolor{black}{Mueller} polarimeter with two high-quality linear polarizers (LPs) and four voltage-controlled liquid crystal variable retarders (LCVRs) for the polarization state generator (PSG) and analyzer (PSA), and a large-area light source suitable for vertical and oblique illumination (indicated in green). An exemplary angular illumination segment for ComSLI is displayed on the LED panel. (\textbf{b}) Vervet brain sample. A 60\,µm coronal section (section no. 506) from the center of a vervet monkey brain. The measured areas include the corpus callosum (cc, mostly in-plane fibers), the cingulum (cg, nearly orthogonal fiber bundles), the corona radiata (cr, fiber crossings), and the fornix (fx, inclined fibers). (\textbf{c}) Human brain sample. A 50\,µm sagittal section (section no. 454) from the human brain that includes the Broca's region. Three investigated regions are indicated in the overview transmittance map. White matter (wm) appears darker than gray matter (gm). }
	\label{fig:setup_and_samples}
\end{figure}

\textcolor{black}{Correlative} 3D-PLI and ComSLI measurements were performed with the Scattering Polarimeter \textcolor{black}{(Fig.\ \ref{fig:setup_and_samples}a)} on several regions of \textcolor{black}{two different samples}: \textcolor{black}{(1) Vervet (Fig.\ \ref{fig:setup_and_samples}b)} -- 60\,µm coronal section from a vervet monkey brain (male, 2.4 years old, no neurological diseases) that contains the corpus callosum, the corona radiata, the cingulum, and the fornix. \textcolor{black}{(2) Human (Fig.\ \ref{fig:setup_and_samples}c)} -- 50\,µm sagittal section from a body donor (80 years old, female, no neurological diseases) that includes Broca's region which is associated with language processing~\cite{Friederici2003}, containing areas of gray and white matter. The resulting parameter maps were compared to those obtained from existing, state-of-the-art 3D-PLI and ComSLI setups \textcolor{black}{(see Methods)}. With these parameter maps, a multimodal fiber direction map was generated by combining data from 3D-PLI and ComSLI. This map displays the most reliable fiber direction signal per pixel from the available modalities, based on a classification algorithm that determines the expected fiber architecture for each image pixel. The multimodal fiber direction map exploits the robustness of 3D-PLI and the ability to detect fiber crossings of ComSLI, increases the total number of evaluated pixels and displays up to three fiber directions. A detailed description of the Scattering Polarimeter and the investigated brain sections is given in the Methods. A comprehensive overview of the fiber direction maps obtained with 3D-PLI \textcolor{black}{and ComSLI} is provided in Supplementary Fig.\ 1.

\subsection*{Three-Dimensional Polarized Light Imaging (3D-PLI)}
\begin{figure}[htbp]
	\centering
	\includegraphics[width=0.8\linewidth]{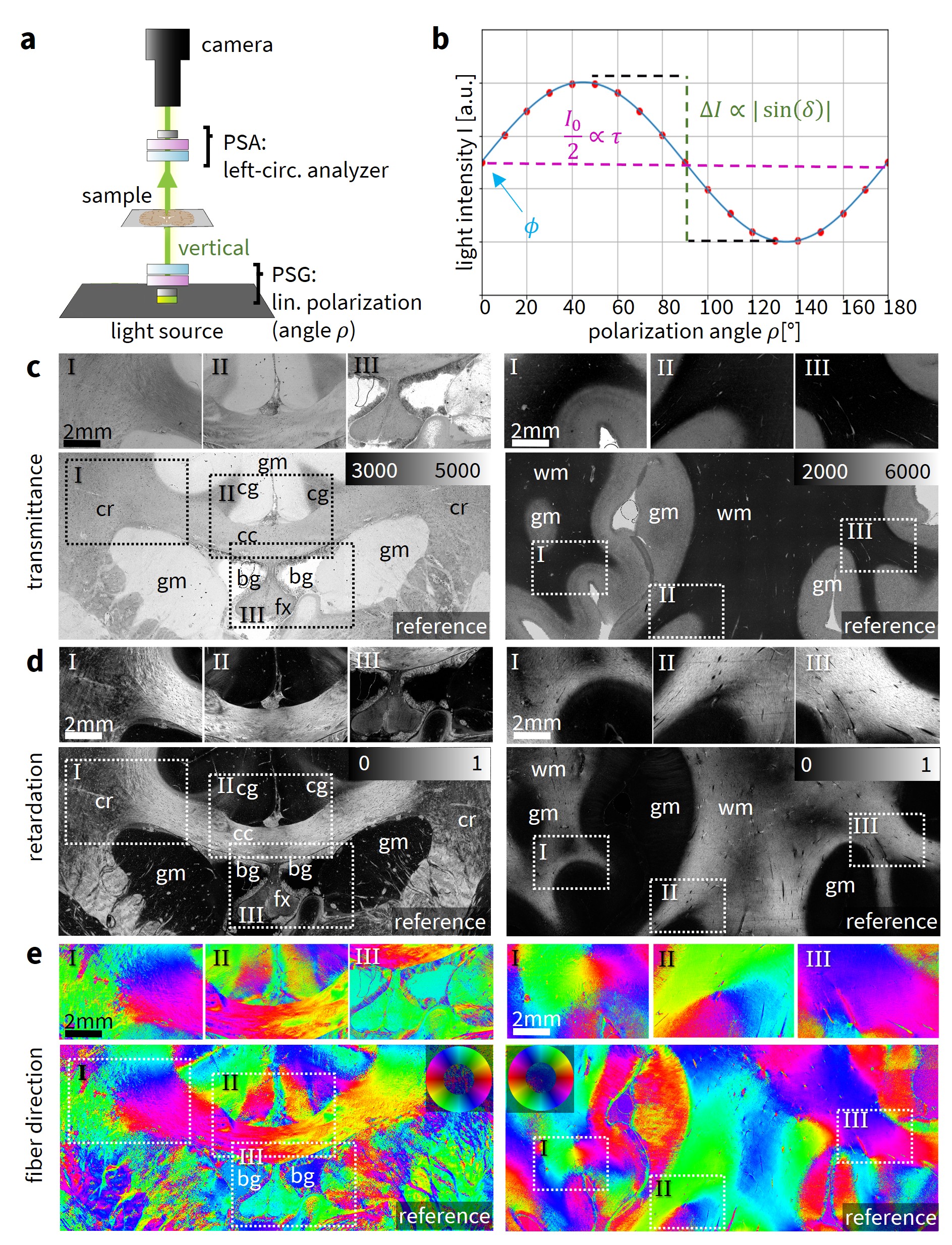}
	\caption{3D-PLI measurement and parameter maps. (\textbf{a}) The Scattering Polarimeter when performing 3D-PLI measurements. The sample is illuminated vertically while the polarization state generator (PSG) generates linear polarization in equidistant azimuthal angles $\rho$ and the polarization state analyzer (PSA) operates as a left-handed circular analyzer. (\textbf{b}) Exemplary signal in 3D-PLI for one image pixel \textcolor{black}{with horizontally oriented fibers ($\phi = 0^{\circ}$)}. Each image pixels yields a sinusoidal profile described by the transmittance $\tau=I_0/2$, the retardation as the normalized amplitude $|\sin(\delta)|=\Delta I/I_0$, and the fiber direction as the phase $\phi$. \textcolor{black}{Image inspired by \textsc{Axer} et al.~\cite{Axer2011a}.} (\textbf{c}) 3D-PLI transmittance maps for the vervet (left) and human (right) brain section compared to a reference measurement (below) with a state-of-the-art 3D-PLI setup \textcolor{black}{(LMP3D)}. The transmittance maps are contrast-matched, indicated structures are the cingulum (cg), corpus callosum (cc), corona radiata (cr), fornix (fx), gray matter (gm), white matter (wm), background (bg). (\textbf{d}) 3D-PLI retardation maps from the Scattering Polarimeter. The inclination map is not shown because it provides the same information as the retardation map. Reference measurement below. (\textbf{e}) Fiber direction maps from the Scattering Polarimeter. The fiber direction is encoded in the hue according to the depicted color wheel. Reference measurement below, corrected with a global phase offset to match azimuthal orientations.}
	\label{fig:pli-results}
\end{figure}

3D-PLI measurements with the Scattering Polarimeter were performed for three regions of the vervet and human brain section, respectively. The polarization state generator (PSG) generated linear polarization angles in steps of $\Delta\rho=10$° while the polarization state analyzer (PSA) operated as a left-handed circular analyzer, as shown in Fig.\,\ref{fig:pli-results}a. From the resulting sinusoidal intensity profile for each image pixel with the properties $I_0$, $\Delta I$ and $\phi$ (as shown exemplary in Fig.\,\ref{fig:pli-results}b), the local transmittance $\tau=I_0/2$ (with signal average $I_0$), retardation $|\sin(\delta)|=\Delta I/I_0$ (signal amplitude $\Delta I$, normalized with $I_0$), and fiber direction $\phi$ (signal phase) \textcolor{black}{were} obtained using a Fourier coefficient fit (see Supplementary\,\textcolor{black}{2} for more details). The retardation was normalized with the retardation of in-plane parallel fibers~\cite{Axer2011}. The reference measurements were performed with the LMP3D (\textit{TAORAD GmbH, Germany}), a state-of-the-art polarization microscope, as explained in the Methods. 

In 3D-PLI, the transmittance represents the average intensity of light transmitted through the tissue during the measurement. The transmittance map has arbitrary units that depend on factors like the brightness of the light source, the camera, and the exposure time. For visual comparison, the minimum and maximum values of the visualization range were chosen so that the reference transmittance map resembles the contrast of the transmittance maps obtained with the Scattering Polarimeter. The direction maps from the LMP3D were rotated by flipping and/or transposing the data array so that the image orientation matches the images from the Scattering Polarimeter. If required, a global direction offset $\phi_{\textrm{off}}$ was added to match the relative fiber directions $\phi_{\textrm{PLI}}=\phi_{\textrm{off}}+\phi'_{\textrm{PLI}}$. 

Figure \ref{fig:pli-results}c--e shows the transmittance maps, the retardation maps (directly related to the fiber inclination), and the color-coded fiber direction maps of the vervet (left) and human (right) brain sections for the Scattering Polarimeter (top) and for a reference measurement with a single-mode, state-of-the-art 3D-PLI setup (bottom). The out-of-plane fiber inclination is not shown in the fiber direction maps in order to simplify the visual comparison of fiber directions in regions with low retardation when judging the performance, i.e.\,for inclined fibers or in gray matter. 

\subsubsection*{Vervet brain section (high transmittance, low scattering)} Unlike the human brain section, the vervet brain section has already been embedded several years ago, leading to a higher transparency of the tissue and hence to a higher transmittance. The transmittance exhibits contrasts between gray matter, white matter, and the background that are comparable to the reference measurement. However, the comparison with the reference measurement demonstrates a transmittance asymmetry that is an artifact of the Scattering Polarimeter measurement. It appears more striking in the vervet than in the human brain section: The lower contrast between gray and white matter (gm/wm) in the vervet brain section -- due to the longer time since embedding of the sample -- requires an adequate choice of the visualized intensity range which, however, visually enhances the asymmetry.
The retardation map agrees with the reference measurement. As expected, the cingulum (cg) and the fornix (fx) show a lower retardation than the corpus callosum (cc). Minor retardation differences, such as an area of slightly lower retardation in the center of the corpus callosum -- indicated by a darker shade of gray -- are displayed correctly. The fiber direction map aligns with the reference measurement for the corpus callosum and the cingulum. Even in the highly complex fibers of the corona radiata (cr), the Scattering Polarimeter yields the same predominant fiber directions as the reference measurement. Fiber directions in white matter are determined with great success. Correct fiber directions in gray matter are detected well for many pixels but not as reliable in comparison. Similar to the Scattering Polarimeter, the LMP3D shows a residual retardation in areas without tissue, indicated by the blue background color in the fiber direction map. The influence of the residual retardation on gray matter fibers can be well observed in region II where the sinusoidal signals are shifted towards the phase of the residual retardation, i.e.\ the blue background direction.

\subsubsection*{Human brain section (low transmittance, high scattering)} The human brain section exhibits excellent contrast between gray and white matter in the transmittance map, aligning well with the reference measurement. Unlike the vervet brain section, the human brain section does not show visible asymmetries in the transmittance map. The retardation map agrees with the reference measurement and displays even minor local phenomena in white matter but also in gray matter, especially close to the white matter border where low retardation fiber bundles can be discerned. Fiber directions are accurately identified in white matter and, for many areas, in gray matter as well. Given the significantly lower fiber density in gray matter, these fibers typically produce a low birefringence signal. 

\subsubsection*{Summary} 3D-PLI measurements with the Scattering Polarimeter show an excellent performance for the human brain sample. Even fibers in regions with low fiber density such as gray matter are resolved correctly. 3D-PLI measurements are usually performed on recently embedded tissue and benefit from a lower transmittance, consequently yielding a good contrast as observed in Broca's region. However, for older, highly transmissive samples like the vervet brain section, the quality of the transmittance map is limited, though fiber directions and retardation remain largely accurate. \textcolor{black}{Mueller} matrix calculus suggests that the asymmetry observed in the vervet sample arises from minor ellipticities in the liquid crystal variable retarders (LCVRs), which have a more pronounced effect on this low-contrast sample compared to the human brain section. The propagation of systematic errors caused by LCVR\,2, 3, and 4 is discussed in Supplementary\,\textcolor{black}{4}.


\subsection*{Computational Scattered Light Imaging (ComSLI)}
\begin{figure}[htbp]
	\centering
	\includegraphics[width=0.9\linewidth]{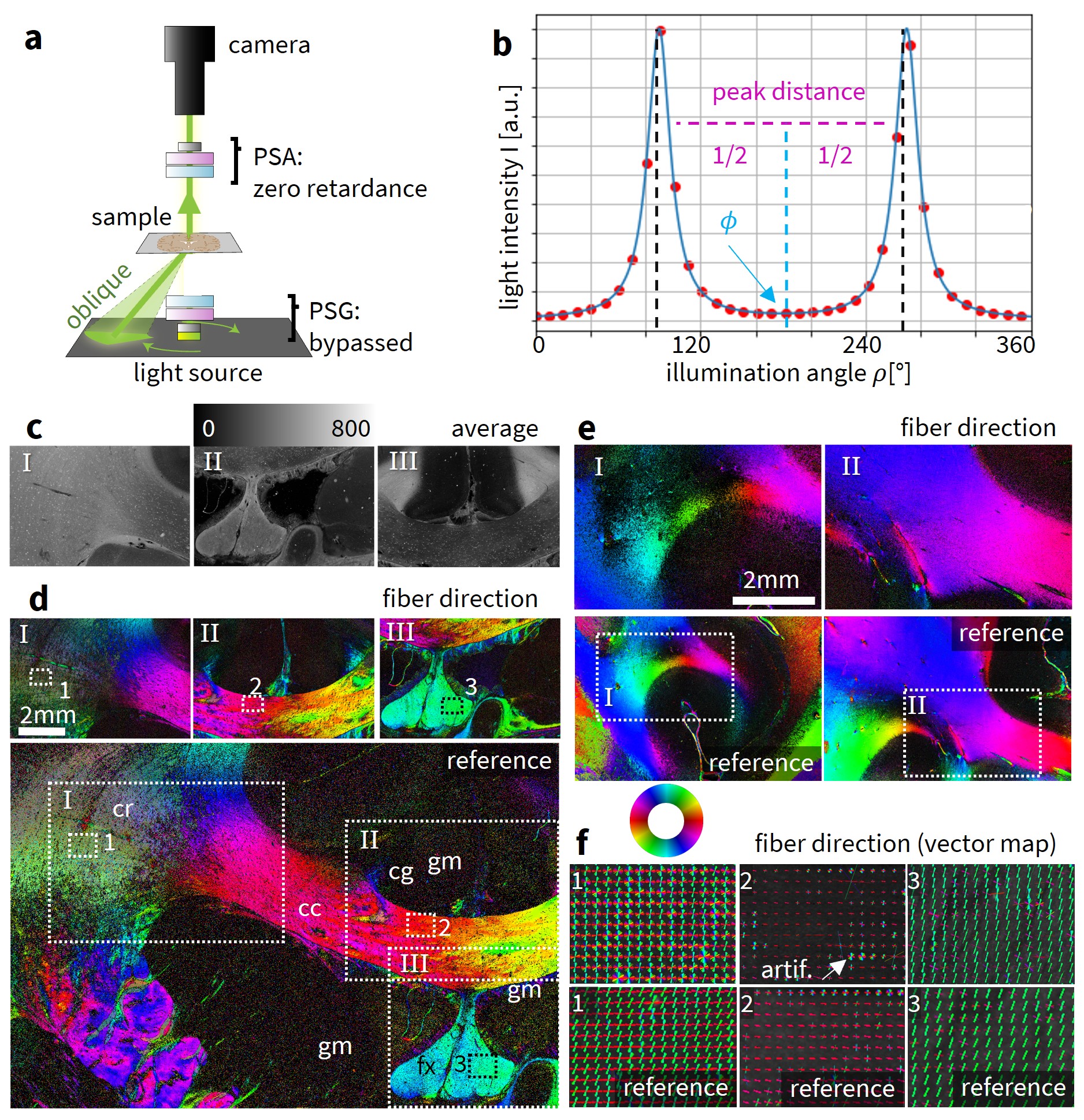}
	\caption{ComSLI \textcolor{black}{measurement and parameter maps}. (\textbf{a}) The Scattering Polarimeter when performing ComSLI measurements. The elements of the PSG are bypassed by the light when illuminating from various angles $\rho$. Only the vertically scattered light is measured.  Even though polarization effects are minor in ComSLI, the elements of the PSA are set to a retardance of 0. (\textbf{b}) Exemplary signal in ComSLI for one image pixel \textcolor{black}{with horizontally oriented fibers ($\phi = 0^{\circ}$)}. For each image pixel, a line profile $I(\rho)$ is obtained. The mid-position of a peak pair is directly related to the fiber direction $\phi$. One peak pair corresponds to parallel fibers, two or three peak pairs correspond to two or three crossing fiber bundles. The peak distance holds information about the fiber inclination. In-plane fibers have a maximum peak distance of 180°. (\textbf{c}) \textcolor{black}{Average scattering signal for three selected regions of the vervet brain section.} (\textbf{d}) \textcolor{black}{Corresponding fiber direction map for the vervet brain sample}, reference measurement with state-of-the-art ComSLI setup below where the regions measured with the Scattering Polarimeter are indicated. Smaller rectangles marked with 1, 2, 3 indicate the regions for which the vector maps are visualized \textcolor{black}{in f}. Corpus callosum (cc), corona radiata (cr), cingulum (cg), fornix (fx), gray matter (gm). (\textbf{e}) \textcolor{black}{Fiber direction maps of two selected regions in the human brain section (regions I and III, cf.\ Fig.\ \ref{fig:setup_and_samples}c)}, reference measurement with state-of-the-art ComSLI setup below where the regions measured with the Scattering Polarimeter are indicated.  (\textbf{f}) Vector maps for the vervet brain sample obtained from the indicated regions \textcolor{black}{in d} with the Scattering Polarimeter \textcolor{black}{(top); kernels of 20 $\times$ 20 pixels are shown as overlaid vectors for better visibility}. Reference vector maps \textcolor{black}{(bottom) with overlaid vectors in kernels of 10 $\times$ 10 pixels}. }
	\label{fig:sli-results}
\end{figure}

Right after the respective 3D-PLI measurements, correlative ComSLI measurements with the Scattering Polarimeter were performed for both samples. The PSG was bypassed by the oblique illumination and the PSA was set to zero retardance, as shown in Fig.\,\ref{fig:sli-results}a. \textcolor{black}{The sample was illuminated under oblique incidence (circular segments on LED panel) with constant polar illumination angle and 15° azimuthal steps (see Methods). For each angle of illumination $\rho$, an image was taken by the camera. Each image pixel in the resulting image series yields an azimuthal intensity profile $I(\rho)$ which is characterized by distinct peaks;} the mid-position of peak pairs is directly related to the fiber directions, as depicted in Fig.\,\ref{fig:sli-results}b. Image data was processed with the Scattered Light Imaging Toolbox (SLIX)~\cite{Reuter2020}. \textcolor{black}{SLIX analyzes the positions of prominent peaks in the intensity profile, and computes and visualizes the in-plane nerve fiber orientations either as a \textit{fiber direction map} or a \textit{vector map}: In the color-coded fiber direction map, each image pixel is represented by 2$\times$2 color-coded sub-pixels that contain all fiber directions within that image pixel. The color-coded vector map shows fiber orientations as vectors whose lengths are weighed with the average scattering map. The vectors are overlaid for $n \times n$ pixels for depiction purposes.}

Figure \ref{fig:sli-results}c depicts the average scattering signal measured with the Scattering Polarimeter. Figure \ref{fig:sli-results}d compares the color-coded fiber direction maps of the vervet brain section with the reference measurement from a state-of-the-art ComSLI setup~\cite{Menzel2021}. Figure \ref{fig:sli-results}e shows the fiber direction maps for two exemplary regions from the human brain section compared to the reference measurement. Figure \ref{fig:sli-results}f shows the vector maps for the areas indicated in d. All displayed vector lengths in the vector maps were weighed with the corresponding average \textcolor{black}{scattering signal}. Vector maps for the Scattering Polarimeter are displayed \textcolor{black}{with overlaid vectors in kernels of 20 $\times$ 20 pixels}, vector maps from the ComSLI setup \textcolor{black}{show overlaid vectors in kernels of 10 $\times$ 10 pixels} which depended on the camera sensor pixels and the objective lens and was chosen to approximately match the vector maps from the Scattering Polarimeter.

\subsubsection*{Vervet brain section (high transmittance, low scattering)} Overall, the Scattering Polarimeter provides fiber directions and crossings with only minor deviations from the reference measurement. Fiber crossings in the corona radiata are resolved similarly to those in the ComSLI setup. The boundary between parallel fibers in the corpus callosum (bold colors) and crossing fibers in the corona radiata (pastel colors, caused by the 2$\times$2 visualization pixels with different colors in crossing regions) follows the same path as in the reference measurement, indicating similar resolution for fiber crossings in both setups.
As expected, tissue borders are significantly influenced by artifacts that scatter light chaotically. Background noise levels are similar in both setups. Fiber directions in the gray matter and the cingulum are not resolved in either the reference measurement or the Scattering Polarimeter measurement: The fiber density in gray matter is very low and the fiber bundles of the cingulum are nearly orthogonal to the sectioning plane which is why the directional signal is low compared to statistical noise. The required exposure time for a clear signal is much longer for the Scattering Polarimeter (4 seconds vs.\,500\,ms). A generally lower signal-to-noise ratio for the Scattering Polarimeter is indicated by the presence of more dark pixels in the fiber direction map.
Vector maps for exemplary subregions suggest that the signal-to-noise ratio is slightly worse for the Scattering Polarimeter, especially for the crossing fibers in the corona radiata. The two crossing fiber directions in the exemplary region are detected but appear noisier (i.e.\,containing false directions, mostly visualized as yellow or blue vectors overlaying the original two directions) in the Scattering Polarimeter. The in-plane parallel fibers of the corpus callosum are detected with comparable precision. Local artifacts—most likely from the embedding medium—manifest as rainbow-colored "wheels" and are independent of the setup, although in the white matter, these artifacts are typically limited to single pixels. The inclined parallel fibers of the fornix show a dominant direction (blue/green) plus a minor secondary direction (red/magenta) for both the ComSLI setup and the Scattering Polarimeter.

\subsubsection*{Human brain section (low transmittance, high scattering)} Increased scattering in the Broca region results from the shorter time since tissue embedding, leading to a lower signal-to-noise ratio that prevents the fiber directions in some white matter areas from being detected compared to the reference measurement. Conversely, \textcolor{black}{region II} exhibits an area with improved pixel evaluation in the indicated tissue regions. Therefore, performance varies depending on the underlying tissue. 

\subsubsection*{Summary} Despite long exposure times, both the Scattering Polarimeter and the state-of-the-art ComSLI setup currently reach their limits for the human brain section, with the state-of-the-art setup performing only slightly better. Both setups yield almost no signal in gray matter. The light source in the Scattering Polarimeter is less bright than in the ComSLI setup, where a brighter light source contributes to better signal quality. The Scattering Polarimeter performs best for the vervet brain section due to the high transmittance in the sample. However, the Scattering Polarimeter suffers from noise in the directional signals, indicated by more black pixels in the fiber direction maps and additional fiber directions or only statistical noise in the vector maps, especially in crossing regions. Detecting fiber directions in gray matter remains challenging for both setups. ComSLI measurements are typically performed on tissue that has been embedded long enough for transmittance to decrease. For relatively freshly embedded samples, both the Scattering Polarimeter and the ComSLI setup reach their limits. High light source brightness and long exposure times are necessary for obtaining a sufficient signal. 


\subsection*{Multimodal fiber direction map}

A multimodal fiber direction map combines information from multiple modalities to achieve the most reliable fiber direction per pixel. The reliability of signals from 3D-PLI or ComSLI largely depends on sample properties like the species, the age (i.e.\,time since embedding), the thickness of the sample~\cite{Woods2011}, and the tissue preparation, but also on the local fiber architecture. The construction of a multimodal fiber direction map was based on a previous classification of image pixels based on the local white matter fiber architecture, as described in the following:

\subsubsection*{Pixel classification (white matter)}
White and gray matter in the human brain section can be well distinguished e.g.\,\textcolor{black}{by their transmittance values} (3D-PLI) in Fig.\,\ref{fig:pli-results} due to the higher scattering and lower transmittance of white matter~\cite{Yaroslavsky2002}. In the vervet brain section, information from both the average map from ComSLI and the retardation map from 3D-PLI is required.

The condition to identify background pixel areas were set as follows: The threshold on the average map from ComSLI was chosen as $\bar{I}\leq20$\,a.u.\ because background areas do not scatter light. Then, the following conditions were used to identify gray matter pixels: 1) The pixel was not already identified as part of the background. 2) The threshold on the retardation map from 3D-PLI was chosen to be $|\sin(\delta)|<0.07$. All areas that are neither background nor gray matter were automatically identified as white matter without the need for additional information. Due to the characteristics of the vervet brain section, the classification is less accurate compared to the human brain section. Nevertheless, major tissue regions were identified correctly. Figure \ref{fig:fom_assembly}a shows the identified white matter pixels in the corona radiata region of the vervet brain sample.

\subsubsection*{Pixel classification (crossing fibers)}
Due to the previously discussed suboptimal signal-to-noise ratio in ComSLI, the most obvious criterion for the identification of fiber crossings -- whether one or multiple fiber directions are found in a pixel -- is currently not sufficient. However, multimodal parameter analysis can still find reliable directional information based on other optical parameters such as average scattering and peak prominence, despite the shortcomings of single individual parameters maps:

Parallel fibers -- regardless of their inclination $\alpha$ (i.e.\, the out-of-plane angle) -- were identified by the following criteria: 1) They were part of the previously identified white matter regions. 2) They had either a) an inclination $\alpha\leq30^\circ$, i.e. a (normalized) retardation $|\sin(\delta)|\geq0.68$ (in-plane fibers) or b) an inclination in the range $30^\circ<\alpha<65^\circ$, i.e. a retardation in the range $0.17<|\sin(\delta)|<0.68$ (inclined fibers), combined with a comparatively high peak prominence, here $\geq 0.3$ for the human brain section and $\geq 1.0$ for the vervet brain section or c) an inclination $\alpha\geq65^\circ$ which corresponds to a retardation $|\sin(\delta)|\leq0.17$ (steep fibers), combined with a strong average scattering intensity of $\bar{I}>1200$\,a.u. for the human brain section and $\bar{I}>430$\,a.u. for the vervet brain section.

Subsequently, fiber crossings, such as in the corona radiata, were identified with the following criterion: They were part of the previously identified white matter regions but not of the already identified parallel fiber structures. Figure \ref{fig:fom_assembly}b shows the pixels that were identified to contain fiber crossings in the vervet corona radiata.

\subsubsection*{Construction of a multimodal fiber direction map}
The construction of a multimodal fiber direction map was based on the pixel classification for different fiber architectures and on the obtained fiber direction in 3D-PLI, ComSLI, and their difference (Fig.\,\ref{fig:fom_assembly}c). Fiber directions from 3D-PLI were selected over those from \textcolor{black}{Mueller} polarimetry due to the superior signal-to-noise ratio, particularly in cortical regions. The fiber directions for all investigated samples are shown in Supplementary Fig.\ 1. The multimodal fiber direction map for the vervet corona radiata is shown in Fig.\,\ref{fig:fom_assembly}d.

Figure \ref{fig:fom_assembly}e illustrates the step-wise procedure used to construct the multimodal fiber direction map. 

The three direction maps of the multimodal fiber direction map were assembled as follows:

\begin{figure}[htbp]
	\centering
	\setlength{\tabcolsep}{0pt}
	\includegraphics[width=0.8\linewidth]{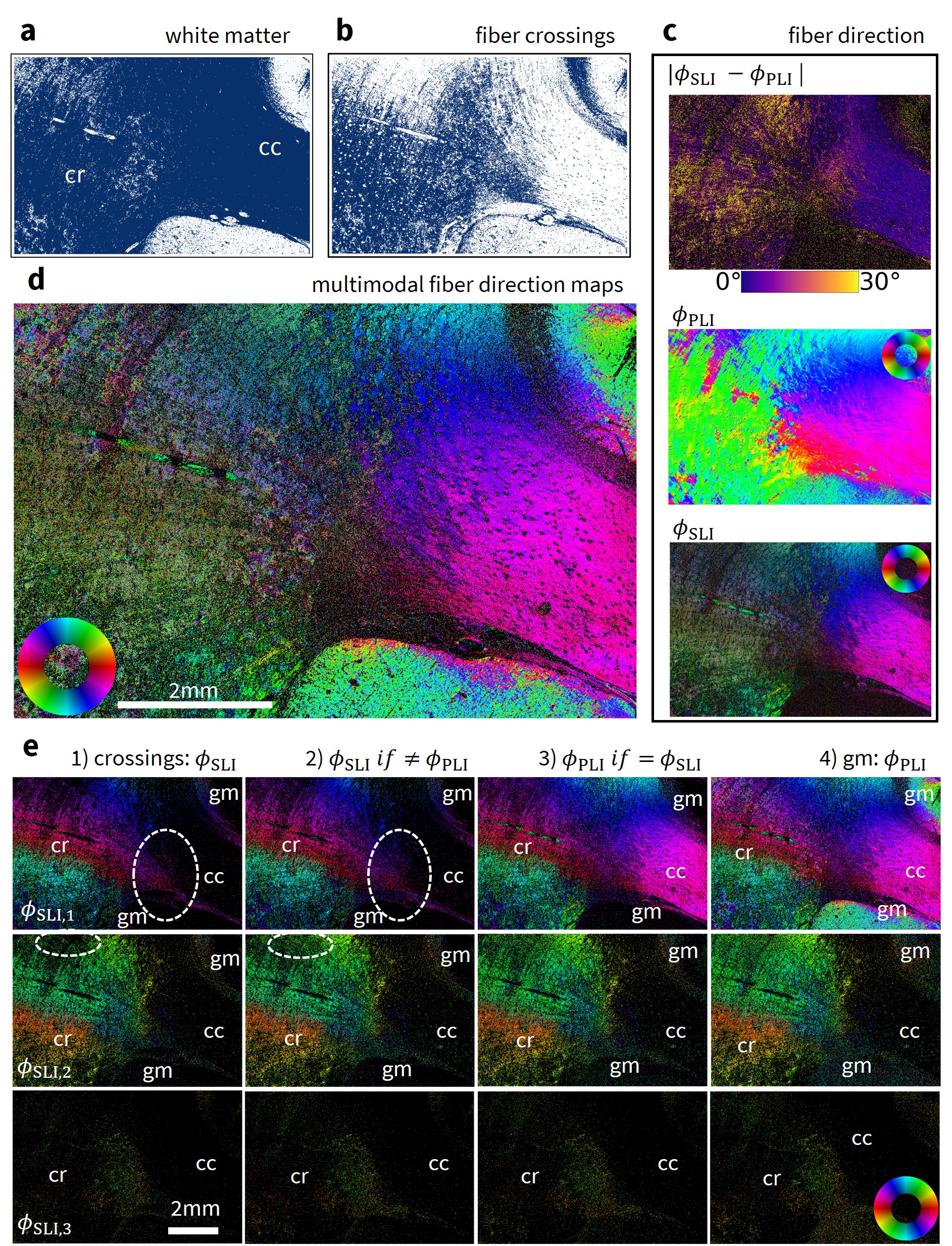}
	\caption{Generation of the multimodal fiber direction map (shown exemplary for the corona radiate region of the vervet monkey brain section). (\textbf{a}) White matter classification. Pixels classified as white matter are shown in blue, pixels classified as gray matter in white. (\textbf{b}) Fiber crossings classification. As a subset of white matter pixels, pixels identified as crossing fibers are shown in blue, other pixels in white. (\textbf{c}) Fiber direction map from 3D-PLI (middle) and ComSLI (bottom) measured with the Scattering Polarimeter; the absolute difference between the direction from 3D-PLI and the closest direction (if there is more than one) obtained with ComSLI is shown at the top. (\textbf{d}) Multimodal fiber direction map. (\textbf{e}) Step-by-step construction of the multimodal fiber direction map: The image pixels are filled based on the pixel classification with the directions $\phi_{\textrm{PLI}}$ from 3D-PLI and $\phi_{\textrm{SLI,1}}$, $\phi_{\textrm{SLI,2}}$, $\phi_{\textrm{SLI,3}}$ from ComSLI. The construction of each multimodal direction map based on the best signal is displayed in three rows. Major changes between the steps are highlighted by dashed ellipses in the image. Corona radiata (cr), corpus callosum (cc), gray matter (gm).}
	\label{fig:fom_assembly}
\end{figure}

\begin{enumerate}
	\item Fiber directions in crossing regions are found more reliably with ComSLI because 3D-PLI cannot differentiate multiple directions and only detects the average signal. For pixels that were identified as part of crossing regions, the directions $\phi_{\textrm{SLI,1}}$, $\phi_{\textrm{SLI,2}}$, and $\phi_{\textrm{SLI,3}}$ were chosen for the respective multimodal direction map, e.g.\,in the corona radiata.
	\item For each pixel within a white matter region not classified as a crossing region, the direction $\phi_{\textrm{SLI,1}}$, $\phi_{\textrm{SLI,2}}$ , or $\phi_{\textrm{SLI,3}}$ was selected when the 3D-PLI and a ComSLI direction map indicate different fiber orientations. This takes into account potential misclassification of border regions that may have been incorrectly identified as non-crossing. Furthermore, it can take into account changes in the optical properties of brain tissue (e.g.\,caused by neurodegenerative diseases~\cite{Wong2014, Novikova2022, Schucht2020, Lee2016}). Here, changes may occur e.g.\ in the birefringence that influence the 3D-PLI signal but not the ComSLI signal. 
	\item For parallel fibers, the sinusoidal curve of the 3D-PLI signal is typically more stable than the line profiles of the ComSLI signal. Additionally, the angular discretization is lower for 3D-PLI than for ComSLI.  For these pixels, $\phi_{\textrm{PLI}}$ was chosen if it was similar to $\phi_{\textrm{SLI,i}}$ with $i\in[1,2,3]$. In this step, changes in the second and third multimodal direction map were small and limited to statistical fluctuations.
	\item Gray matter was treated separately due to its rather low signal which is usually detected more reliably by 3D-PLI than ComSLI. For the first direction, $\phi_{\textrm{PLI}}$ was chosen when $\phi_{\textrm{SLI,1}}$ was not found. Otherwise, $\phi_{\textrm{SLI,1}}$ was chosen. For the second and third direction, $\phi_{\textrm{SLI,2}}$ and $\phi_{\textrm{SLI,3}}$ were used for the respective direction map, if found. Only a few pixels exhibited three distinct fiber directions, leading to an overall darker appearance in the third directional map.
\end{enumerate}
The three new multimodal direction maps were assembled into a single, multi-colored fiber direction map, where each image pixel is represented by 2$\times2$ color-coded subpixels. If one pixel contains two or three fiber directions, the 2x2 subpixels show two or three colors, respectively. All pixels without any detected fiber direction are displayed in black. They are mostly part of the background area or gray matter. 

\subsubsection*{Comparison between single-mode and multimodal fiber direction maps} 
\begin{figure}[htbp]
	\includegraphics[width=\linewidth]{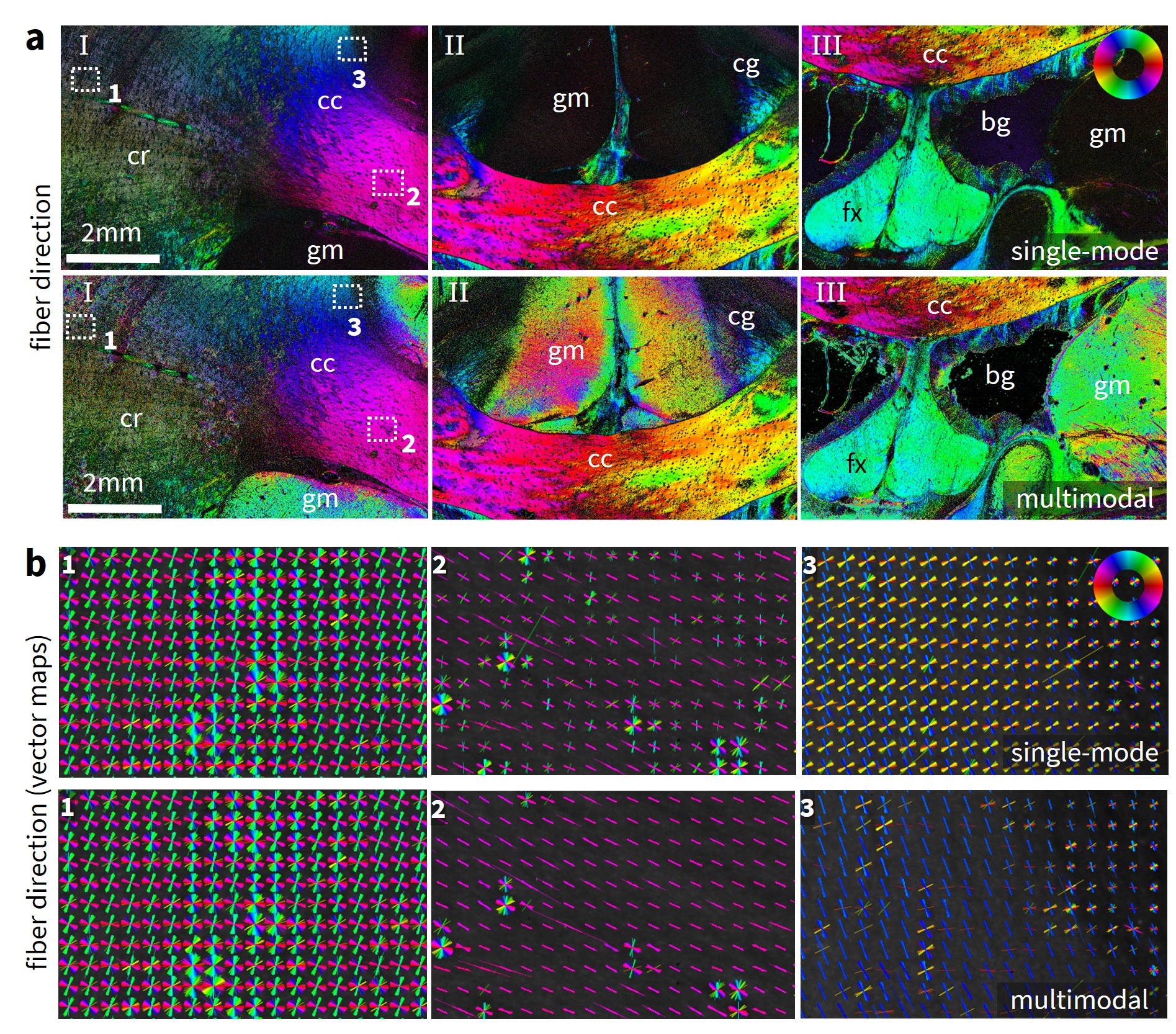}
	\caption{Comparison between single-mode and multimodal fiber direction maps for the vervet brain section. (\textbf{a}) Single-mode fiber direction maps obtained from a ComSLI measurement with the Scattering Polarimeter. Corona radiata (cr), corpus callosum (cc), cingulum (cg), fornix (fx), gray matter (gm), background (bg). Below, the corresponding multimodal fiber direction maps obtained from a combination with the corresponding 3D-PLI measurement. Compared to the single-mode fiber direction map obtained only with ComSLI (first row), the multimodal fiber direction map combines up to three fiber directions per pixel and exhibits a noticeably better signal-to-noise ratio. Background (bg) noise is inherently filtered out. (\textbf{b}) Single-mode vector maps for three selected regions indicated  with 1, 2, 3 in a. The displayed vector length is weighed with the corresponding average \textcolor{black}{scattering signal} and vectors are visualized \textcolor{black}{in kernels of 20 $\times$ 20 pixels}. Below, the corresponding multimodal vector maps with same visualization. The maps show more subtle differences between the single-mode directions obtained only with ComSLI and the multimodal directions: Fiber directions are more distinct in the multimodal vector map and less affected by noise that leads to additional fiber directions in 2 and 3 for parallel fibers in the single-mode vector map. Crossing fibers as in 1 do not change significantly because they are mostly retrieved from ComSLI alone.}
	\label{fig:fom_comparison}\end{figure}

Figure \ref{fig:fom_comparison} compares the single-mode fiber direction maps obtained from ComSLI measurements with the Scattering Polarimeter, and the multimodal fiber direction maps obtained from a combination with the 3D-PLI fiber direction maps for the vervet brain sample. The single-mode and the multimodal vector maps (vector length weighed with the corresponding average map and vectors \textcolor{black}{overlaid for kernels of 20 $\times$ 20 pixels}) are compared for three selected areas indicated in the fiber direction maps. The multimodal fiber direction maps do not only provide information about multiple fiber directions obtained from ComSLI but also exhibit a better signal-to-noise ratio due the data from 3D-PLI, particularly for parallel fibers and gray matter, compared to the single-mode  fiber direction maps obtained only with ComSLI. Noticeable improvements are observed in the gray matter, the corpus callosum, and parts of the cingulum. Additionally, background noise is inherently filtered out by the routine.

More subtle differences are evident in the vector maps: While the crossing regions in 1 do not differ significantly between the single-mode and the multimodal  fiber direction maps (as expected, as they are all retrieved from the ComSLI directions), the directions of the parallel in-plane fibers in the corpus callosum indicated by 2 are more distinct and less affected by noise. Although a second green direction manifests in the single-mode fiber direction maps, the multimodal fiber direction maps show only the dominant magenta direction for most pixels. The parallel fibers of the corpus callosum area bordering the gray matter in 3 are improved even more visibly, nearly removing the additional yellow direction. Overall, the multimodal fiber direction maps provide more detailed information with improved quality.

However, a notable phenomenon occurs in the cingulum. Here, where the fiber bundles run nearly orthogonal to the sectioning plane, many pixels remain black in the multimodal fiber direction maps, even when 3D-PLI clearly detects a dominant direction. This occurs because ComSLI fails to reliably detect a single or multiple directions. As a result, these pixels are not filled with 3D-PLI data because the 3D-PLI fiber directions are not considered reliable based on information from ComSLI. Previous studies discussed how the inclination obtained from 3D-PLI is not necessarily reliable for very steep fibers and how a combination of ComSLI and 3D-PLI can improve the results in this context~\cite{Menzel2021a}.


\section*{Discussion}
Scattering polarimetry seeks to optimally combine \textcolor{black}{3D-PLI and ComSLI} to make use of the strengths and compensate the weaknesses of each modality. All modalities can benefit from a brighter light source, a more sensitive camera and a thorough parameter study to determine optimal measurement settings for each modality. 

The Scattering Polarimeter has successfully advanced from the prototypical stage to a fully operational multimodal device. The novel device integrates oblique illumination for ComSLI and vertical illumination for 3D-PLI into a single, multimodal system that has maintained the flexibility to realize illumination with different wavelengths or the use of other polarization states \textcolor{black}{to perform full Mueller polarimetry}. Software for measurement and evaluation were developed. 
\textcolor{black}{The current study focused on enabling combined measurements of 3D-PLI and ComSLI in a single device and is a first proof-of-concept study. Future studies are needed to further optimize the measurement routine and enable high-throughput measurements of many tissue samples.}
The most urgent hardware upgrade is a brighter LED light source. Especially the ComSLI reference measurements indicated that measurements with a brighter light source lead to a better signal and less noise. The development of a custom light source with integrated polarization state generator (PSG) is optional but must be considered when scatterometry ComSLI~\cite{Menzel2021} or compressed sensing ComSLI~\cite{Heiden2024} become part of the routine measurement. For high-speed measurement applications, the integration of a custom scanning table is mandatory. 
\textcolor{black}{In its current implementation, the 3D-PLI and ComSLI measurements with the Scattering Polarimeter take considerably more time than their respective reference measurements, partly because of more optical elements in the light path. Future studies should look into how the measurement time can be further reduced, e.g., by reducing the number of images (the LMP3D uses 9 instead of 18 images) and reducing the exposure time when implementing a brighter light source and a more sensitive camera. Even if measurements with the Scattering Polarimeter take longer than for state-of-the-art setups, there is no extra time needed for exchanging tissue samples or for performing image registration, which significantly reduces the total time needed for correlative measurements and analysis.}

Scattering polarimetry was demonstrated for samples with notably different optical properties. The performance was investigated using exemplary tissue samples with different scattering and transmission properties. Sources of statistical and systematic errors were identified and suggestions for further improvement were made. The human brain sample most closely resembles common measurement conditions: In 3D-PLI, samples are measured days after embedding. In ComSLI, it is beneficial to wait several weeks or even months until the transmittance has increased but it is uncommon to measure a sample that was embedded years earlier like the vervet brain sample. The spectrum of tissue samples revealed the strengths and limitations of each measurement modality, and has highlighted potential challenges that can arise with different tissue properties. A parameter study could now identify the optimal measurement parameters for routine scattering polarimetry of common tissue samples. The primary goal would be to determine the default exposure time and required number of repetitions for all modalities to obtain a good signal-to-noise ratio. Recently embedded tissue required relatively longer exposure times to yield a sufficient signal, especially for ComSLI. The optimal time for a multimodal measurement after embedding should be investigated through a long-term study. In the following, the key findings for each modality are discussed.

Even though \textcolor{black}{retardation and fiber direction could in principle be retrieved from Mueller matrix elements with Lu-Chipman decomposition (see Supplementary \textcolor{black}{1}), the 3D-PLI analysis used here (Supplementary \textcolor{black}{2})} provides the most reliable measurements of fiber orientations. The Fourier fit effectively smooths statistical noise and stabilizes the signal, allowing the detection of weak signals even in gray matter, where the fiber density is comparatively low. However, \textcolor{black}{this method} has a limitation due to the lack of retardation calibration which becomes apparent for low contrast samples such as the vervet brain sample. Any elliptical deviation introduced by the system can distort the transmittance and, to a lesser extent, the retardation. This issue is inherent to the method but more apparent for the Scattering Polarimeter because the LCVRs are less precise than rotating mechanical elements. The Scattering Polarimeter showed an excellent performance for the human brain sample. Even fibers in regions with low fiber density, such as gray matter, were correctly resolved. For highly transmitting older samples such as the vervet brain sample, the limits of the Scattering Polarimeter became visible in the transmittance map, even though fiber directions and retardation in white matter were still detected correctly. Comparison with the reference measurement shows excellent agreement for the human brain sample. Results for the highly transmitting vervet brain sample deviate slightly from the reference measurements due to an asymmetry in the transmittance; however, fiber directions and retardation are still determined accurately, apart from minor deviations in some gray matter areas where fiber density is low.

ComSLI is the only method among the available modalities capable of measuring multiple fiber directions within a single image pixel \textcolor{black}{and} offers a higher resolution compared to techniques such as dMRI. The average \textcolor{black}{scattering} map generated by ComSLI is highly reliable for tissue classification for both older and newer tissue. However, ComSLI has a lower signal-to-noise ratio compared to 3D-PLI, which is particularly noticeable in gray matter and more recently embedded (i.e.\,less transmittive and more scattering) tissue. The signal-to-noise ratio can be enhanced for the Scattering Polarimeter by implementing a brighter light source, larger illumination segments, a more sensitive camera and -- if time is not a factor -- using a longer exposure time and more repetitions. Good light-shielding is crucial because reflections negatively influence the signal. However, ComSLI measurements successfully identified the major directions in agreement with the reference measurement even though the signal quality was lower compared to the reference measurement, which utilized a brighter light source and a high-sensitivity camera. Fiber crossings are sometimes misidentified as statistical noise and therefore not evaluated, e.g.\,in gray matter but also in the cingulum. Conversely, additional fiber directions are incorrectly identified in parallel fibers due to statistical noise. While dominant directions are already well recognized, there is a need for more accurate determination of fiber crossings for each pixel. To address these challenges, it is essential to improve the hardware to achieve less noisy measurements. As a second step, adequate signal interpolation can aid with extracting correct fiber directions. With the corresponding enhancements, ComSLI measurements with the Scattering Polarimeter are expected to improve significantly in the future.

\textcolor{black}{Mueller} polarimetry, despite \textcolor{black}{not being investigated in the main part of this study}, can determine all the optical properties that 3D-PLI measures. Unlike 3D-PLI, \textcolor{black}{Mueller} polarimetry can also distinguish between linear and circular retardation, as well as measuring the diattenuation which could be used to better distinguish between tissue compositions~\cite{Menzel2017, Menzel2019}. Moreover, \textcolor{black}{Mueller} polarimetry measures depolarization, which can be used to differentiate between white and gray matter and between healthy and tumorous (disordered, chaotically growing) fibers. The Eigenvalue Calibration Method (see Supplementary 1) provides a theoretically ideal calibration routine, with the primary limiting factor being the quality of calibration samples. However, \textcolor{black}{Mueller} polarimetry requires a good signal-to-noise ratio, i.e.\,a long exposure time, a bright light source, a high sensitivity camera and sufficient repetitions because each \textcolor{black}{Mueller} coefficient is determined by a sum of only four images. \textcolor{black}{First measurements (Supplementary 1 and 3) show the potential of performing full Mueller polarimetry with the Scattering Polarimeter, but more studies are needed to further optimize its performance, especially when extracting fiber orientations. The fiber orientations reconstructed by Mueller polarimetry are highly consistent with those derived from 3D-PLI analysis, but Mueller polarimetry fails to reconstruct fiber orientations in the highly scattering human brain sample and in gray matter regions (see Supplementary Fig.\ 1, on the left).}

The field of multimodal parameter analysis is extensive, with the Scattering Polarimeter generating more than 30 different maps (around 15 for \textcolor{black}{Mueller} polarimetry, 4 for 3D-PLI, and 12 from SLIX for ComSLI), covering a wide range of optical parameters, not even including the three-dimensional visualization of fiber orientations and vector maps. This work provides a proof-of-principle of how multimodal parameter analysis serves the imaging of nerve fiber architecture. It is meant to showcase the potential of scattering polarimetry and multimodal parameter analysis. Generally, multimodal parameter analysis would not necessarily require a Scattering Polarimeter when good image registration of different modalities is available. However, correlative measurements with pixel-precise alignment of images are very beneficial for resolving structures of nerve cells that occur at this level of detail and allow for much faster multimodal measurement and analysis. 

Accurate tissue and fiber classification is crucial for multimodal parameter analysis. Automatic approaches such as histogram-based brightness analysis or machine learning-based methods should be explored to generalize the classification to all sorts of unknown samples. A routine to assemble a multimodal fiber direction map was developed, based on the established fiber direction map from ComSLI. When all measurement techniques are optimized and tissue and fiber classification routines are enhanced, the multimodal fiber direction map is expected to yield even more valuable results.\\

In this study, the Scattering Polarimeter was introduced, enabling correlative measurements between 3D-PLI, ComSLI, and potentially \textcolor{black}{Mueller} polarimetry due to its design as a \textcolor{black}{Mueller} polarimeter based on liquid crystal variable retarders. This allows for pixelwise mapping and cross-validation of fiber orientations in brain samples, leveraging the strengths of each technique. The performance of the device was evaluated using two exemplary brain sections with different transmitting and scattering properties, one more suitable for 3D-PLI and one more suitable for ComSLI, thus highlighting both the strengths and challenges of each measurement modality.
Error analysis based on \textcolor{black}{Mueller} matrix calculus gave further insight into the imaging properties of the device. Suggestions for further improvements were made, such as the implementation of a brighter light source. As a proof of concept, a multimodal fiber direction map was created, combining the advantages of both techniques. 

A possible application for correlative measurements and multimodal parameter analysis is the investigation of neurodegenerative diseases: When myelin degenerates -- often in the course of neurodegenerative diseases~\cite{Wong2014, Novikova2022, Schucht2020, Lee2016} -- the consequent birefringence change leads to an apparent change of fiber directions: The existence of myelin leads to a strong negative birefringence. When myelin degenerates, only the remaining axon exhibits a weaker and positive birefringence. However, the fiber directions obtained by ComSLI are not influenced by this phenomenon because scattering is mostly unaffected by the effect~\cite{Georgiadis2024}. Multimodal analysis can support the investigation of neurodegenerative diseases by comparing ComSLI and 3D-PLI directions. 

Although the Scattering Polarimeter is still in its early developmental stages, the ability to perform correlative measurements of 3D-PLI, ComSLI, and \textcolor{black}{potentially Mueller} polarimetry has significant potential for neuroimaging applications and could become a valuable tool in the future.


\section*{Methods}
\label{methods_and_materials}
\paragraph{Tissue preparation and samples.}
All samples were cryo-sectioned. First, the brains were removed from the skull within 24 hours after death. Subsequently, they underwent fixation using a buffered 4\% formaldehyde solution to prevent decay and were stored at room temperature for several weeks (vervet brain) or months (human brain). Subsequently, the brains were deep-frozen for storage and sectioning, thus requiring previous cryo-protection~\cite{Pegg2002}: The formation of ice crystals in the tissue was prevented by immersing the vervet brain in a solution of 10\% gly\-cerin and then 20\% gly\-cerin (each for several days) while being stored in the fridge~\cite{Rosene1986}. The human brain was immersed in a solution of 20\% gly\-cerin with phosphate-buffered saline (PBS) and 0.5\% sodium azide for several days and stored at room temperature. Afterwards, each brain was dipped in isopentane at room temperature for some minutes and frozen in -80$^\circ$C isopentane. A large-scale cryostat microtome (\textit{CM3600 (Leica)} for the human brain, \textit{CM3500 (Leica)} for the vervet brain) was used to cut the frozen brain into thin sections. A blockface image of the brain block surface was taken before each cut to support image registration~\cite{Axer2011a}. The brain sections were mounted onto glass slides, embedded in 20\% glycerin, cover-slipped, weighted, and -- after air bubbles had emerged from the solution -- sealed with nail polish. 

The investigated human and vervet brain tissue samples contain gray and white matter, distinct anatomical regions with characteristic fiber structures (in-plane parallel fibers, out-of-plane fibers, crossing fibers, but also artifacts such as crystallization of the embedding medium and bubbles within the medium). The transmittance of cryo-sectioned brain tissue changes over time because scatte\-ring within the tissue decreases due to evaporation of the embedding medium. Therefore, polarization-based measurements are usually done within days after the preparation process. Scattering-based measurements can actually benefit from the decreased transmittance due to a higher contrast of scattering peaks~\cite{Menzel2019}. Accordingly, scattering-based measurements of cryo-sectioned tissue are ideally performed several weeks or even months after their preparation. Here, the correlative measurements were performed right after each other. Two different samples were investigated:

The human brain sample is a 50\,µm sagittal section from a human brain (female, 80 years, no neurological diseases) that includes the Broca's region and areas of gray and white matter (see Fig.\,\ref{fig:setup_and_samples}c). The brain was obtained in accordance with the ethics committee at the Heinrich Heine University Düsseldorf, Germany, through the body donor program of the Institute of Anatomy I, University Hospital of Heinrich Heine University Düsseldorf, Germany. The ethics committee confirmed that such postmortem human brain studies do not require any additional approval if a written informed brain donation consent of the subject or their next of kin or legal representative is available. For the human brain used in this study, such a consent is available. The study was performed in accordance with the rules of the local ethics committee (study number 2023-2632). The section was embedded in 2023 (less than a year old at the time of measurement), making it very new compared to the other sample. There are no crystallization artifacts. The transmittance is comparatively low while the scattering is high.

The vervet brain sample is a 60\,µm coronal section from a vervet monkey brain (male, 2.4 years old, no neurological diseases). Euthanasia procedures conformed to the AVMA Guidelines for the Euthanasia of Animals, using ketamine/pentobarbital anesthesia followed by perfusion with phosphate buffered saline and fixation with 4\% paraformaldehyde. All animal procedures and experimental protocols were approved by the Wake Forest Institutional Animal Care and Use Committee (IACUC \#A11-219), in accordance with the National Institutes of Health guidelines for the use and care of laboratory animals, and in compliance with the ARRIVE guidelines. The selected section contains the corona radiata, the fornix, the cingulum, and the corpus callosum, but also gray matter in the cortex (see Fig.\,\ref{fig:setup_and_samples}b). The section was embedded in 2012 (i.e.\,12 years before the measurements) and the embedding medium has started to crystallize at the tissue borders. Compared to more recently embedded samples, the transmittance is higher. The sample contains a variety of representative fiber configurations in close proximity that fit well in the field of view of the Scattering Polarimeter. Three regions were measured that together contain the corpus callosum (in-plane fibers), the cingulum (nearly orthogonal fiber bundles), the border region between corona radiata (crossing fibers) and corpus callosum (in-plane fibers), the fornix (inclined fibers), and areas with gray matter or background. 

\paragraph{Setup of the Scattering Polarimeter.}
The design of the Scattering Polarimeter is shown in Fig.\,\ref{fig:setup_and_samples}a. The two linear polarization filters (LP, \textit{Thorlabs LPVISC100}) are marked in gray. The LPs have an extinction ratio of 10000:1 in the range from 520\,nm to 740\,nm. The voltage-controlled liquid crystal variable retarders (LCVRs, \textit{Thorlabs LCC1223-A}) are displayed in pink and blue, depending on the azimuthal orientation of their slow axis (45° or 0°, respectively). The LCVRs with a 20\,mm large clear aperture are antireflection (AR)-coated to work best in a wavelength range from 350\,nm to 700\,nm. In LCVRs, the retardance $\delta = \delta(V)$ depends on the applied voltage $V$ in a non-linear way. The retardance-voltage curve was determined in a crossed-polarizer configuration for all four LCVRs.

LP\,1 has its transmission axis at an azimuthal angle of 0$^\circ$. LCVR\,1 has its slow axis at 45$^\circ$ and LCVR\,2 at 0$^\circ$. Together, the LP\,1, LCVR\,1 and LCVR\,2 form the \textit{polarization state generator} (PSG). The \textit{polarization state analyzer} (PSA) mirrors the optic components of the PSG: Here, LCVR\,3 with its slow axis at 0$^\circ$ is followed by LCVR\,4 with its slow axis at 45$^\circ$ that again is followed by LP\,2 with its transmission axis at 0$^\circ$. The tissue sample is located between the PSG and the PSA. With this configuration, the PSG can generate any polarization state, and the PSA can analyze any polarization state. The LCVRs can operate as a classic \textcolor{black}{Mueller} polarimeter but also perform 3D-PLI measurements by generating linear polarization with equidistant angles $\rho$ and using the PSA as a circular analyzer for left- and right-circular polarization.

The \textit{Basler acA5472-17uc} CCD camera (\textit{Sony IMX183 CMOS} sensor with a size of 13.1$\times$8.8\,mm², a sensor pixel size of 2.4$\times$2.4\,µm$^2$ and 5472$\times$3648 pixels) was combined with the \textit{QIOPTIQ APO-RODAGON-D 1X 75/4,0} objective lens to meet the geometrical restrictions of the setup: The imaging area needed to be as large as possible to provide a good spatial overview over the sample while the field of view was constrained by the apertures of the optical elements, especially the PSG. The free working distance of the objective lens needed to accommodate the elements of the PSA. Overall, this results in a field of view of 8$\times$5\,mm$^2$\textcolor{black}{, a pixel size in object space of about 1.5\,µm, and an optical resolution of 2.19\,µm}  (determined with a United States Air Force (USAF) chart). The camera aperture was set to its maximum value of 5.6 for maximum light incidence.

Two light paths are indicated in Fig.\,\ref{fig:setup_and_samples}a in green with the color representing the green illumination wavelength: The vertical light path for \textcolor{black}{Mueller} polarimetry and 3D-PLI, and an exemplary angular segment of oblique illumination for ComSLI. A large-area light source with separately controllable LEDs (\textit{INFiLED s1.8 LE Indoor LED Cabinet}) was implemented to display ComSLI illumination patterns but also illuminate the sample directly from below. The panel consists of 256$\times$256 LEDs (R, G, B) with an LED pitch of 1.8\,mm and a view angle of 120$^\circ$. It has a size of 46.56$\times$46.56\,cm². The overall brightness is 1000 cd/m². The response of the LCVRs depends on the wavelength, thus the light source spectrum was narrowed down by a matching spectral filter (\textit{Thorlabs FBH515-10} with a central wavelength of $\lambda_0=514.5$\,nm and a full width half maximum (FWHM) of 10\,nm) positioned on top of the light source in the vertical light path. 
\textcolor{black}{The sample was positioned at a height of 16\,cm above the LED panel, in focus distance of the camera.}

\paragraph{Measurements with the Scattering Polarimeter.}
The illumination sequence started with a central circle with the diameter of roughly the PSG apertures that serves as the light source for vertical illumination (in \textcolor{black}{Mueller} polarimetry and 3D-PLI) while the voltage-controlled PSG and PSA are set to the corresponding retardance to achieve the required polarization states. 
\textcolor{black}{The Mueller polarimetry measurements were performed as described in Supplementary 1.} For 3D-PLI, the PSG needs to generate linear polarization at equidistant azimuthal steps $\rho$ within the range of 180$^\circ$ while the PSA operates as a circular analyzer. For this purpose, the voltage of LCVR\,1 was varied while LCVR\,2 operated as a quarter-wave plate, thus turning the elliptical polarization generated by LCVR\,1 into linear polarization with an azimuthal angle $\rho$ that is determined by the retardance $\delta_1$ of LCVR\,1. LCVR\,3 and LCVR\,4 were removed from the setup. Then, $\rho(\delta_1)$ was measured in a crossed-polarizer configuration. \textcolor{black}{For this study, steps of $\rho=10^{\circ}$ were chosen for 3D-PLI, resulting in 18 images.}

After the vertical illumination and the measurement of 3D-PLI and \textcolor{black}{Mueller} polarimetry, the angular illumination segments for ComSLI illumination follow one after each other. The angular segments are defined by their inner and outer radius (in pixels), the number of segments, the azimuthal segment size and their color (green as a default). 
For ComSLI, green illumination patterns with 15° angular steps (i.e.\ 24 measurement angles), a segment width of 12°, an inner radius of 50 LEDs, and an outer radius of 90 LEDs were used\textcolor{black}{, corresponding to a polar illumination angle of about $30-45^{\circ}$.} This illumination sequence has comparatively large illumination segments for maximum brightness while maintaining a sufficiently precise angular discretization.
In ComSLI, LCVR\,1 and LCVR\,2 are bypassed by the oblique light incidence. LCVR\,3 and LCVR\,4 are set to a retardance of $\delta_3=\delta_4=0$ to exclude any possible polarization dependency of the scattered light.

The exposure times for polarimetric measurements was chosen to be 2 seconds, the exposure time for scattering-based measurements was 4 seconds, short enough to avoid overexposure and long enough for a sufficient signal-to-noise ratio. No gain was used. Only the green camera channel was evaluated. All images were measured four times and averaged.

\paragraph{Reference measurements.}
Reference measurements were performed with state-of-the-art setups specialized on single-mode measurements for 3D-PLI and ComSLI, respectively. For 3D-PLI, the LMP3D (\textit{Taorad GmbH, Germany}) setup was used. \textcolor{black}{In contrast to the Scattering Polarimeter which employs fixed liquid crystal retarders and polarizers, the LMP3D contains mechanically rotating optical elements: a rotating linear polarizer in front of the sample and a fixed circular analyzer (consisting of a quarter-wave plate ($\lambda=532$\,nm) followed by another linear polarizer) behind the sample.} The microscope setup employs the \textit{SVS Vistek evo4070 GigE} CCD camera with a sensor dimension of 2048$\times$2048 pixels combined with a \textit{Nikon 4x (NA 0.2)} lens, thus achieving \textcolor{black}{a pixel size in object space of 1.85\,µm, and an optical resolution of 1.94\,\textmu m}. The light source has a wavelength of $\lambda=(520\pm20)$\,nm and is combined with a wavelength filter with $\lambda=(532\pm5)$\,nm. The first polarization filter is rotated to equidistant angular positions with $\Delta\rho=20^\circ$\textcolor{black}{. One 3D-PLI measurement consists of 9 images with 10\,ms exposure time each, and takes about 2 seconds in total (including mechanical rotation).} A motorized XY stage scans the imaging area in tiles of 3.8$\times$3.8\,mm$^2$. 

The ComSLI setup \textcolor{black}{ is comparable to the setup shown in Fig.\ \ref{fig:sli-results}a, without PSG and PSA. It} contains the \textit{Basler acA5472-17uc} CCD camera which is located centrally above the sample holder. The camera employs a \textit{Sony IMX183 CMOS} sensor with a sensor dimension of 13.1$\times$8.8\,mm$^2$, a sensor pixel size of 2.4$\times$2.4\,µm$^2$ and 5472$\times$3648 pixels. It can operate at a maximum of 17 frames per second and provides either a color depth of 10 bit for 10-bit high-speed all-pixel readout or a color depth of 12 bit for 12-bit high-resolution readout. In this configuration, the \textit{Rodenstock Apo-Rodagon-D120} objective lens yields a field of view of 1.6$\times$1.1\,cm$^2$\textcolor{black}{, a pixel size in object space of about 2.9\,µm, and an optical resolution of 3.5\,µm} at a focal length of 120\,mm. Illumination is provided by a controllable LED panel, specifically the \textit{Absen Polaris 3.9pro In/Outdoor LED Cabinet}, featuring 128$\times$128 LEDs (R, G, B) with an LED pitch of 3.9\,mm and an overall brightness of 5000\,cd/m$^2$. The LED panel measures 50$\times$50\,cm$^2$. A sample holder is positioned 17\,cm above the light source\textcolor{black}{.} The dimensions of the Scattering Polarimeter are based on these sizes. \textcolor{black}{ The ComSLI setup uses green angular illumination with circle segments in steps of 15° with a segment width of 9°. For the vervet brain, an inner radius of 27 LEDs and an outer radius of 36 LEDs were used as segment length, corresponding to a polar illumination angle of $32-40^{\circ}$. For the human brain, larger segments were chosen (inner radius of 22 LEDs and outer radius of 55 LEDs) to provide brighter illumination for the less transmissive sample.} The vervet brain section was measured with an exposure time of 500\,ms, a gain of 3\,dB and 4 repetitions. The human brain section was measured with an exposure time of 5\,seconds, a gain of 0\,dB and 4 repetitions, using the \textit{SVS-VISTEK HR455CXGE} CCD camera. 

\textcolor{black}{\paragraph{Data analysis.} 
	Mueller polarimetry measurements were analyzed as described in Supplementary 1. 
	3D-PLI signals were analyzed as shown in Fig.\ \ref{fig:pli-results}b and explained in Supplementary 2: A discrete Fourier analysis was applied to the sinusoidal intensity curve of each measured image pixel to compute transmittance (related to the signal average), retardation (related to the signal amplitude), and in-plane fiber direction (related to the signal phase).
	ComSLI signals were analyzed as shown in Fig.\ \ref{fig:sli-results}b: For each measured image pixel, the prominent peaks in the resulting intensity profile were determined with the Scattered Light Imaging ToolboX (SLIX) \cite{Reuter2020,Menzel2021a} and used to compute the in-plane fiber directions per pixel: For one prominent peak, the fiber direction was computed from the  position of the peak itself, for two prominent peaks, from the mid-position of the peak pair, and for more than one peak pair, from the mid-position of each pair (if the peaks lie $180^{\circ} \pm 35^{\circ}$ apart). All fiber direction angles are in degrees, with 0$^\circ$ being along the positive x-axis and 90$^\circ$ along the positive y-axis.
}


\section*{Data availability}
All data generated and analyzed during this study are available in the Jülich DATA repository (\url{https://doi.org/10.26165/JUELICH-DATA/CUXQYF}).

\bibliography{main}

\section*{Acknowledgements}
We thank Markus Cremer and the laboratory team at Forschungzentrum Jülich (INM-1), Germany, for preparing the brain sections, Philipp Schlömer (INM-1) for the 3D-PLI measurements with the LPM3D and support with hardware choice, and Roger Woods from the UCLA Brain Research Institute for providing the vervet brain sample (National Institutes of Health under Grant Agreements no.\ R01MH092311 and 5P40OD010965).
This work has received funding from the Klaus Tschira Boost Fund of the German Scholars Association, from the Deutsche Forschungsgemeinschaft (DFG) under project no.\,498596755, from the European Union’s Horizon 2020
Research and Innovation Programme grant no.\ 945539 (“Human Brain Project” SGA3) and no.\ 101147319 (“EBRAINS 2.0 Project”). M.M. received funding from the Helmholtz Association’s Initiative and Networking Fund through the Helmholtz International BigBrain Analytics and Learning Laboratory (HIBALL) under the Helmholtz International Lab grant agreement InterLabs-0015. Computing time was granted through VSR Computing Time Projects on the supercomputer JURECA at Jülich Supercomputing Centre (JSC), Germany. The funders played no role in study design, data collection, analysis and interpretation of data, or the writing of this manuscript. 

\section*{Author contributions}
F.H. substantially contributed to the design and construction of the Scattering Polarimeter, the conception and design of the study, and the analysis and interpretation of the experimental data. Furthermore, F.H. created the figures, and wrote the first draft of the manuscript. M.A. contributed to the interpretation of the experimental data, especially the 3D-PLI data, and to the revision of the manuscript. K.A. contributed to the anatomical content of the study and to the revision of the manuscript. M.M. participated in the design of the setup and the study, contributed to the interpretation of the experimental data, especially the ComSLI data, and to the revision of the manuscript. All authors read the final manuscript and gave approval for publication.

\section*{Additional information}
\textbf{Supplementary information} accompanies this paper.
\paragraph{Competing interests:} The authors declare no competing interests.\\


\renewcommand{\figurename}{Supplementary Figure}
\renewcommand\thefigure{\arabic{figure}} 
\setcounter{figure}{0}

\cleardoublepage

\section*{\textcolor{black}{Supplementary 1: Mueller polarimetry}}

\textcolor{black}{The Mueller polarimetry measurements with the Scattering Polarimeter were performed by setting the polarization state generator  and analyzer (PSG and PSA) to all possible combinations of the basic Stokes polarization states (vertical and horizontal linear polarization, diagonal and anti-diagonal linear polarization, right-handed and left-handed circular polarization) and measuring the transmitted intensity for each configuration, resulting in 36 measurements. The required polarization settings were achieved with voltages previously determined for all four liquid crystal variable retarders (LCVRs). Subsequently, the $4 \times 4$ Mueller matrix elements were calculated by solving the resulting system of linear equations~\cite{Ghassemi2009}. } 


\subsection*{\textcolor{black}{Eigenvalue calibration method}}

\textcolor{black}{To account for non-ideal behavior of PSG and PSA, all measured Mueller matrices were calibrated following the eigenvalue calibration method (ECM) described by \textsc{Compain} et al.\cite{Compain1999}. The method determines the calibration matrices $A$ and $W$ from a Mueller polarimetry measurement without sample, a measurement with a linear polarization filter in two different orientations, and a measurement with a retarding element. An advantage of the ECM is that it does not require the azimuthal angle of the calibration samples to be precisely aligned. Here, we used a linear polarization filter (\textit{Edmund Optics XP38}) at azimuthal angles of $\approx$0° and $\approx$45° and a quarter-wave plate (\textit{Newport 10RP34-532}) at $\approx$45°. From the known calibration matrices $A$ and $W$, every measured \textcolor{black}{Mueller} matrix $M_{\textrm{meas}}$ was calibrated via: $M = AM_{\textrm{meas}}W$.} 


\subsection*{\textcolor{black}{Lu-Chipman decomposition}}

\textcolor{black}{To derive the (linear) retardance and the fast axis (fiber) orientation from the measured Mueller matrices, the Lu-Chipman decomposition~\cite{Lu1996} was used. It} rewrites the \textcolor{black}{Mueller} matrix $M$ as the matrix product: 
\begin{equation}
	\label{eq:theory:luchipman}
	M = M_\Delta M_R M_D
\end{equation}
with $M_\Delta$ the depolarization matrix, $M_R$ the retardance matrix and $M_D$ the diattenuation matrix. The decomposition of the retardance and the \textcolor{black}{depolarization} matrix includes solving an eigenvalue problem for the \textcolor{black}{depolarization} matrix~\cite{Lu1996}.\\

When $M$ is decomposed according to Eq.\,(\ref{eq:theory:luchipman}), the \textcolor{black}{Mueller} matrix of depolarization $M_\Delta$ yields the total depolarization $\Delta$:
\begin{equation}
	\Delta = 1-\frac{|Tr(M_{\Delta})|}{3} , \qquad 0\leq \Delta \leq 1
\end{equation}
with $Tr(M_\Delta)$ denoting the trace of the matrix. A value of 1 indicates complete depolarization. A value of 0 indicates that no depolarization occurs. 
\\

\textcolor{black}{The total retardance $R$ can be computed from the retardance matrix $M_R$ via}: 

\begin{equation}
	R = \cos^{-1}\left(\frac{Tr(M_R)}{2}-1\right)
	\qquad \textrm{with} \qquad 
	M_R=\begin{pmatrix}
		1 & \vec{0}^T\\
		\vec{0} & m_R
	\end{pmatrix}
\end{equation}

\textcolor{black}{It} can be separated into an optical rotation \textcolor{black}{and} the linear retardance $\delta$~\cite{Manhas2006, Ghosh2008}:

\begin{equation}
	\delta = \cos^{-1}\left(\sqrt{\left(\left[m_{R,11}+m_{R,22}\right]^2+\left[m_{R,21}+m_{R,12}\right]^2\right)}-1\right) 
\end{equation}
with the orientation of the fast axis with respect to the \textcolor{black}{x-axis (fiber orientation) given by}:
\begin{equation}
	\phi_R = \frac{1}{2}\tan^{-1}\left(\frac{m_{R,12}-m_{R,21}}{m_{R,13}-m_{R,31}}\right).
\end{equation}


\textcolor{black}{\subsection*{Measured Mueller matrices of known optical elements}}

\textcolor{black}{To test the performance of the Scattering Polarimeter as Mueller polarimeter, we performed Mueller polarimetry measurements of test samples with known optical properties and calibrated the measured Mueller matrices with the ECM as described above.} 

The ideal \textcolor{black}{Mueller} matrix for a measurement without sample $M_{\textrm{air, id.}}$ is the unit matrix~\cite{Goldstein2011}. The normalized calibrated \textcolor{black}{Mueller} matrix $M_{\textrm{air}}$ for the empty setup was measured to be: 
\begin{equation}
	M_{\textrm{air}}=
	\begin{pmatrix}
		1.0000&  0.0009 & -0.0017 & -0.0018\\
		-0.0004 &  0.9978 & -0.0023 &-0.0004\\
		0.0000& 0.0027& 0.9971 & 0.0047\\
		-0.0004 & 0.0042& -0.0035&  0.9967\\
	\end{pmatrix} 
	\qquad M_{\textrm{air, id.}}=
	\begin{pmatrix}
		1 &  0 & 0 & 0\\
		0&   1 & 0 &0\\
		0& 0 & 1  & 0 \\
		0 & 0 & 0 & 1\\
	\end{pmatrix} 
\end{equation}

The ideal \textcolor{black}{Mueller} matrix for measuring a horizontal linear polarizer $M_{\textrm{LPH, id.}}$ contains only ones and zeros~\cite{Goldstein2011}. The normalized calibrated \textcolor{black}{Mueller} matrix $M_{\textrm{LPH}}$ for a linear polarization filter (\textit{Edmund Optics XP38}) in horizontal orientation (disregarding the pre-factor of 0.5) was measured to be:
\begin{equation}
	M_{\textrm{LPH}}=
	\begin{pmatrix}
		1.0000 &0.9904 &0.1171 & -0.0001\\
		1.0349 & 0.9910& 0.1192 &0.0068\\
		0.0174 &0.0209& -0.0059 & -0.0312\\
		0.0032& -0.0064 &0.0301& 0.0201
	\end{pmatrix} \qquad M_{\textrm{LPH, id.}}=
	\begin{pmatrix}
		1 &  1 & 0 & 0\\
		1&   1 & 0 &0\\
		0& 0 & 0  & 0 \\
		0 & 0 & 0 & 0\\
	\end{pmatrix} 
\end{equation}

The matrix element $m_{10}=1.0349$ is larger than $m_{00}$. This can happen due to statistical fluctuations, but applying the criteria suggested by \textsc{del Hoyo et al.}~\cite{Hoyo2020} sets the matrix element to $1$ which yields a physically correct \textcolor{black}{Mueller} matrix $M_{\textrm{LPH}}$.

The ideal \textcolor{black}{Mueller} matrix of a birefringent plate $M_{\textrm{QWP', id.}}$ \textcolor{black}{(\textit{Newport 10RP34-532})} was modeled as a phase retarder with $\delta=0.255$, i.e.\,$ \cos(2\cdot\pi\cdot0.255)\approx-0.03$ and $\sin(2\cdot\pi\cdot0.255)\approx1$, hence taking the offset from the ideal quarter-wave plate (QWP) into account by using a phase shift of 0.255 instead of 0.25~\cite{Goldstein2011}. The normalized calibrated \textcolor{black}{Mueller} matrix $M_{\textrm{QWP'}}$ of the birefringent plate was measured as:
\begin{equation}M_{\textrm{QWP'}}\approx
	\begin{pmatrix}
		1.0000  &  0.0203&  0.0131 & 0.0222\\
		-0.0136& -0.0518& 0.1114 &  -0.8284\\
		-0.0014  & 0.055 &  0.8889 & 0.0561\\
		0.0138&  0.8910& -0.0990& -0.0601
	\end{pmatrix}
	\qquad M_{\textrm{QWP', id.}}\approx
	\begin{pmatrix}
		1 &  0 & 0 & 0\\
		0&   -0.03 & 0 &-1\\
		0& 0 & 1  & 0 \\
		0 & 1 & 0 & -0.03\\
	\end{pmatrix} 
\end{equation}

For the empty setup, the matrix elements are in accordance with the theoretical matrix down to the third or even fourth decimal place. For the linear polarization filter,  the matrix elements deviate mostly in the third or second decimal position. For the birefringent plate, the matrix elements mostly deviate in the second decimal place. 
Overall, the performance of the Scattering Polarimeter is comparable to other \textcolor{black}{Mueller} polarimeters in literature~\cite{Bueno2000, Baba2002}. 


\newpage
\section*{\textcolor{black}{Supplementary 2: Signal analysis in Three-dimensional Polarized Light Imaging (3D-PLI)}}

For a 3D-PLI measurement, \textcolor{black}{the sample is illuminated by linearly polarized light under different azimuthal rotation angles $\rho$. The linearly polarized light passes through the sample (thin cryo-sectioned brain tissue) and the optical anisotropy of the myelinated nerve fibers introduces a phase shift depending on the local nerve fiber orientation which changes the polarization to elliptical. To analyze the phase shift, the light passes through a circular analyzer. In the Scattering Polarimeter, polarization state generator and analyzer are realized by a set of liquid crystal variable retarders and fixed linear polarizers (see main Fig.\ 1a). In the state-of-the-art 3D-PLI setup that was used as reference measurement (LMP3D, see Methods), the ingoing linear polarization is generated by a rotating linear polarizer, and the circular analyzer is realized by a quarter-wave plate followed by another linear polarizer}. The pixelwise evaluation \textcolor{black}{of a 3D-PLI signal} is depicted in \textcolor{black}{main Fig.\,2}b. The light intensity $I(\rho)$ of every image pixel is described by \textcolor{black}{a sinusoidal curve\cite{Axer2011}}:
\begin{equation}
	\label{eq:stateoftheart:pli_intensity}
	I(\rho) = \frac{I_0}{2}\cdot (1+\sin(2\rho-2\phi)\cdot\sin(\delta)),
\end{equation}
where $\delta$ is the retardance, $\sin(\delta)$ is the retardation, $\rho$ is the \textcolor{black}{angle of ingoing linear polarization,} and $\phi$ is the fiber direction angle (projected onto the brain section plane with respect to the \textcolor{black}{x-axis} of the setup). The local fiber inclination (i.e., the out-of-plane angle) $\alpha$ is connected to the local retardance:
\begin{align}	
	\delta = \frac{2\pi d}{\lambda} \Delta n \cos^2(\alpha).
\end{align}
The retardation $|\sin(\delta)|$ is linked to the amplitude $\Delta I$ of the intensity curve $I(\rho)$ via: $\Delta I\propto|\sin(\delta)|$. The phase of $I(\rho)$ is determined by the fiber direction (i.e.\,the in-plane angle) $\phi$. 

A discrete harmonic Fourier decomposition allows the evaluation of the curve $I(\rho)$~\cite{Glazer1996}:
\begin{equation}
	I(\rho) = \frac{I_0}{2}\cdot (1+\sin(2\rho-2\phi)\cdot\sin(\delta)) = a_0 + a_2\cdot\sin(2\rho)+b_2\cdot\cos(2\rho)
\end{equation}
with the coefficients
\begin{equation}
	a_0 = \frac{I_0}{2},\qquad a_2=\frac{I_0}{2}\cdot\sin(\delta)\cdot\cos(2\phi),\qquad b_2=\frac{I_0}{2}\cdot\sin(\delta)\cdot\sin(2\phi).
\end{equation}
The Fourier coefficients $a_0$, $a_2$ and $b_2$ are calculated for each pixel from the individual light intensity $I_i$ for all equidistant angles $\rho_i$ for $N$ sampled angular steps $i$:
\begin{equation}
	a_0 = \frac{1}{N}\sum_{N}^{i=1} I_i ,\qquad 
	a_2=\frac{2}{N}\sum_{N}^{i=1} I_i\sin(2\rho_i) ,\qquad 
	b_2=\frac{2}{N}\sum_{N}^{i=1} I_i\cos(2\rho_i).
\end{equation}
From the Fourier coefficients, the transmittance map, the retardation map, and the fiber direction map can be computed.
The \textit{transmittance map} corresponds to the average of the signal and represents the birefringence-independent light extinction of the sample. White and gray matter have distinct attenuation coefficients in the optical regime, with the attenuation through white matter being generally larger due to higher absorption and scattering. The transmittance $\tau$ is calculated pixelwise from the zeroth Fourier coefficient as:
\begin{equation}
	I_0 =\tau = 2a_0. 
\end{equation}
The \textit{retardation map} shows the parameter $|\sin(\delta)|=\Delta I/I_0$, i.e.\,the normalized amplitude of the light intensity profile. The retardation $|\sin(\delta)|$ is computed from the three Fourier coefficients:
\begin{equation}
	|\sin(\delta)| = \frac{\sqrt{a_2^2+b_2^2}}{a_0} \in [0,1].
\end{equation}

The fiber direction map shows the azimuthal (in-plane) angle $\phi_{\textrm{PLI}}$ of each fiber given by:
\begin{equation}	\phi_{\textrm{PLI}} = 180^\circ \cdot \frac{\arctan2(-a_2, b_2)}{2\pi} \in [-90^\circ, 90^\circ]
\end{equation}
and shifted by 90$^\circ$ to be in the angular range $\in [0^\circ, 180^\circ]$.


\newpage
\section*{Supplementary \textcolor{black}{3}: Fiber directions from all modalities}
\label{appendix:direction-comparison}

\begin{figure}[h!]
	\centering	
	\includegraphics[width=\linewidth]{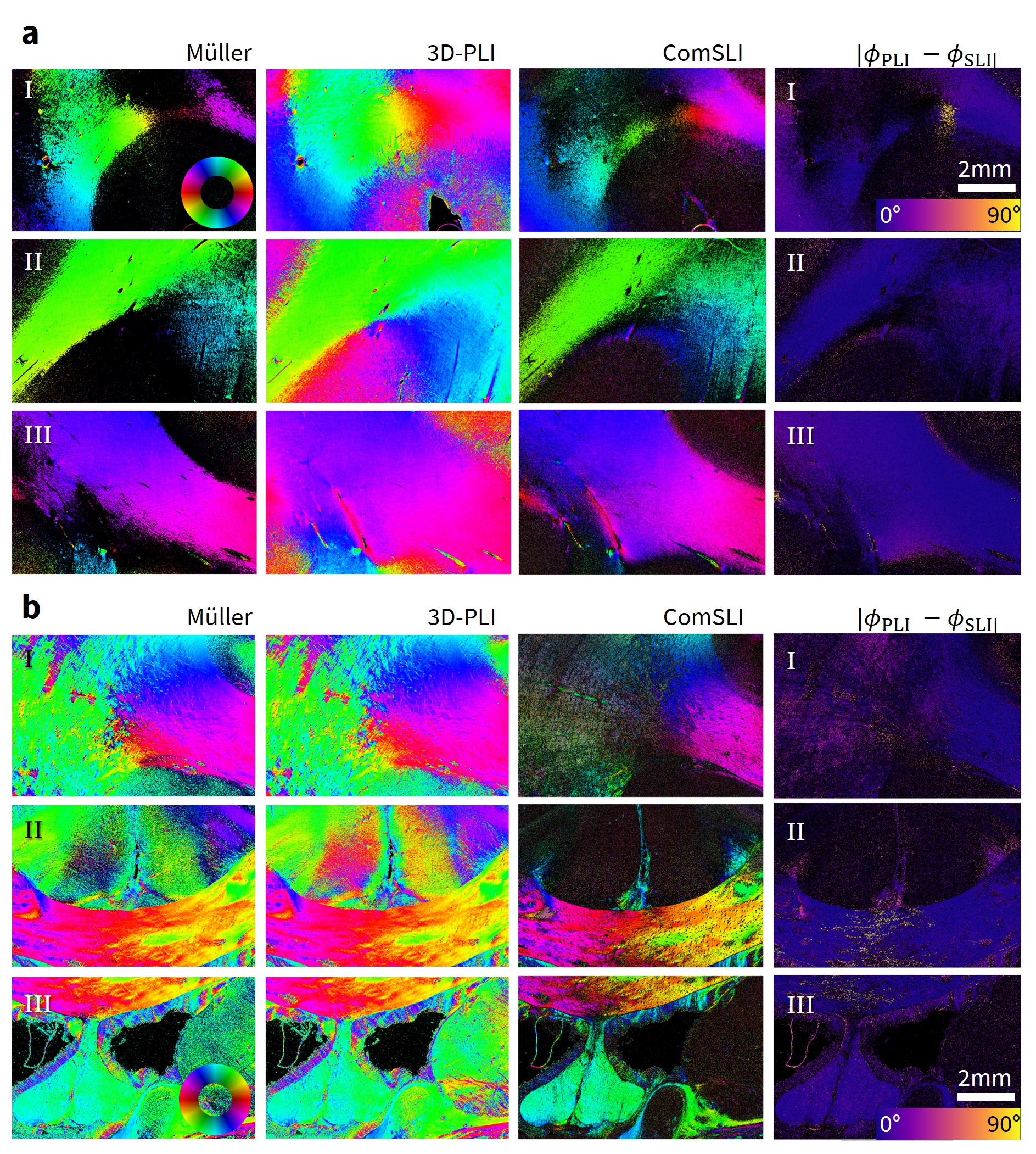}
	\caption{Comparison of fiber directions from all modalities: \textcolor{black}{Mueller} polarimetry, 3D-PLI and ComSLI (all three directions displayed in a fiber direction map). The fourth column shows the absolute difference in fiber direction between 3D-PLI and the closest direction obtained by ComSLI $|\phi_{\textrm{PLI}} - \phi_{\textrm{SLI}}|$. Background areas are displayed in black. (\textbf{a}) Human brain section (high scattering, low transmittance). (\textbf{b}) Vervet brain section (low scattering, high transmittance).}	
	\label{fig:multimodal:direction-comparison}
\end{figure}


\newpage
\section*{Supplementary \textcolor{black}{4}: Calculation of systematic errors}
\label{appendix:errorcalculation}

To uncover statistical and systematic errors within the setup, image pixels from two characteristic regions of the human brain section (low transmittance, high scattering) and the vervet monkey brain section (high transmittance, low scattering) were investigated in detail. 

The insets in \textcolor{black}{Supplementary Figs.\,\ref{fig:3dpli-fit}a and \ref{fig:3dpli-fit}c} show the transmittance map for the human and the vervet brain section, respectively, with two locations marked for each: one in white matter (magenta star), where a sinusoidal curve with high retardation (i.e.\,large amplitude) but comparatively low transmittance is expected. Reference measurements indicated parallel fibers in the chosen white matter area. The phase of the sinusoidal curve depends on the fiber direction but also on the circular analyzer setting of the PSA. The right-handed circular analyzer (RC) setting results in a 90° shift compared to the left-handed circular analyzer (LC). This shift must be accounted for in the correct mathematical evaluation of fiber directions but has no other consequences. Furthermore, a region without tissue (magenta diamond) was investigated, accordingly with high transmittance and only statistical noise, both due to the absence of birefringent tissue.

Figures \ref{fig:3dpli-fit}a and \ref{fig:3dpli-fit}c display the raw intensities averaged over 20$\times$20 image pixels, thus limiting the impact of single pixel artifacts, blood vessels, or crystals within the embedding medium. The statistical error $\sigma_I$=17\,a.u.\,was estimated from the average standard deviation of 4 raw images. The sinusoidal curve is displayed for the right-circular (RC, blue) and left-circular (LC, orange) setting. A dashed line indicates the fitted sinusoidal curve. Figures \ref{fig:3dpli-fit}b and \ref{fig:3dpli-fit}d show the corresponding residuals. The coefficient of determination $R^2$ indicates the fit quality for each image pixel:
\begin{equation}
	R^2 = 1 - \frac{\sum (y_i - f_i)^2}{\sum (y_i - \bar{y})^2} \in[0,1]
\end{equation}

where $y_i$ represents the measured data points (here, intensity in arbitrary units), $f_i$ the fitted data, and $\bar{y}$ the mean value of the measured data. The closer $R^2$ is to 1, the better the fit quality typically is. With a coefficient of determination $R^2$ close to 1, the Fourier coefficient fit can generally be considered successful, especially for white matter areas. 

However, the mean amplitude (i.e.\,transmittance) differs slightly between RC and LC, indicating a polarization dependency of transmittance. This phenomenon is not based on any physical tissue properties as confirmed by the reference measurements but caused by minor ellipticities within the scattering polarimeter, i.e.\,unintended elliptical deviations from the ideal circular or linear polarization.

Furthermore, the residuals indicate a systematic error in the shape of a higher-frequency sinusoidal curve. This systematic error is relatively small compared to the maximum amplitude of the signal in white matter and even smaller for the background signal, suggesting that the higher-frequency curve either originates from the tissue itself or is enhanced through interaction with the tissue. 

\textcolor{black}{Mueller} matrix calculus provides further insight into the observed systematic errors, as demonstrated in the following. Fibrous brain tissue with a retardance $\delta$ and a fiber direction $\phi$ can be modeled as a \textcolor{black}{Mueller} matrix of a retarder rotated around an azimuthal angle $\phi$~\cite{Zhang2003}: 
{\begin{align}
		\label{appendix:eq:tissue-ret}
		&M_{\textrm{tissue}}\\&=R(\phi)\cdot M(\delta)\cdot R(-\phi)\\&=
		\begin{pmatrix}
			1 & 0 & 0 & 0\\
			0 &	\cos(2\phi) & -\sin(2\phi) & 0\\
			0 & \sin(2\phi) & \cos(2\phi) & 0\\
			0 & 0 & 0 &1\\
		\end{pmatrix}
		\begin{pmatrix}
			1 & 0 & 0 & 0\\
			0 & 1 & 0 &0\\
			0 & 0&	\cos\delta & \sin\delta \\
			0 & 0& -\sin\delta & \cos\delta \\
		\end{pmatrix}
		\begin{pmatrix}
			1 & 0 & 0 & 0\\
			0 &	\cos(2\phi) & \sin(2\phi) & 0\\
			0 & -\sin(2\phi)& \cos(2\phi) & 0\\
			0 & 0 & 0 &1\\
		\end{pmatrix}\\&=
		\begin{pmatrix}
			1 & 0 & 0 & 0\\
			0 &	\cos^2(2\phi)+\sin^2(2\phi)\cos\delta & \cos(2\phi)\sin(2\phi)(1-\cos\delta) & -\sin(2\phi)\sin\delta\\
			0 & \cos(2\phi)\sin(2\phi)(1-\cos\delta) & \sin^2(2\phi)+\cos^2(2\phi)\cos\delta & +\cos(2\phi)\sin\delta\\
			0 & +\sin(2\phi)\sin\delta & -\cos(2\phi)\sin\delta &\cos\delta\\
		\end{pmatrix}\\&=
		\begin{pmatrix}
			1 & 0 & 0 & 0\\
			0 &	w^2_\phi+v^2_\phi\cos\delta & w_\phi v_\phi(1-\cos\delta) & -v_\phi\sin\delta\\
			0 & w_\phi v_\phi(1-\cos\delta) &v^2_\phi+	w^2_\phi\cos\delta & +w_\phi\sin\delta\\
			0 & +v_\phi\sin\delta & -w_\phi\sin\delta &\cos\delta\\
		\end{pmatrix}
	\end{align}
	
	Here, the abbreviations $v_\phi=\sin(2\phi)$, $w_\phi=\cos(2\phi)$ are used. Diattenuation and depolarization are neglected.
	
	\textcolor{black}{Diattenuation is neglected because previous studies have shown that the diattenuation of brain tissue does not have much impact on 3D-PLI signals \cite{Menzel2017}.}
	
	\textcolor{black}{Depolarization is neglected because depolarization values of healthy white matter are expected to be relatively low, also in formalin-fixed tissues (\textsc{Gros} et al.\cite{Gros2023} reported values between 0.9--1.0). Furthermore, depolarization effects do not affect the in-plane fiber orientations from 3D-PLI or the transmittance, but only the determined retardance values, which were not analyzed quantitatively in this study. This can be seen as follows:
		Assuming that brain tissue also acts as a non-uniform partial depolarizer characterized by depolarization indices $\Delta_a$, $\Delta_b$, $\Delta_c$, it is possible to extend the mathematical model to incorporate depolarization effects:}
	\begin{align}
		\begin{pmatrix}
			1 & 0 & 0 &0 \\
			0 & \Delta_a & 0 & 0 \\
			0 & 0 & \Delta_b & 0\\
			0 & 0 &0 & \Delta_c \\
		\end{pmatrix} \cdot M_{\textrm{tissue}}
	\end{align}
	\textcolor{black}{Multiplying the resulting matrix with the matrix of a circular analyzer (from the left) and the Stokes vector of ingoing linear polarization with polarization angle $\rho$ (from the right) yields for the intensity of a 3D-PLI measurement:}
	\begin{equation}
		I = \frac{I_0}{2}\large(	1 + \Delta_c\sin(2\rho-2\phi)\sin(\delta)\large)
	\end{equation}
	\textcolor{black}{As can be seen, the amplitude of the signal (retardance) is affected by the depolarization index, but not the phase of the signal (fiber orientation) or the average (transmittance).}
	
	\subsection*{\textcolor{black}{Mueller} calculus for a non-ideal \textcolor{black}{Mueller} polarimeter}
	Ideal and non-ideal \textcolor{black}{Mueller} polarimeters can be modeled using \textcolor{black}{Mueller} matrix calculus. In the following, the impact of non-ideal optical properties (usually arising from the non-ideal optical components) within the polarization state generator (PSG) and/or the polarization state analyzer (PSA) in a non-ideal \textcolor{black}{Mueller} polarimeter is mathematically described. The \textcolor{black}{Mueller} polarimeter is composed of four variable retarders and two linear polarization filters, oriented in the same manner as in the scattering polarimeter. 
	
	With ideal retarders at azimuthal angles $\theta$ and the Stokes vector of horizontal linear polarization $\vec{S}_H$, the Stokes vector $\vec{S}_{\textrm{PSG}}$ for the ideal polarization state generator (PSG) is given by:
	\begin{align}
		&\vec{S}_{\textrm{PSG}}\\ &= M(\delta_2,\theta=0^\circ)\cdot M(\delta_1, \theta=45^\circ)\cdot \vec{S}_H \\&=	\begin{pmatrix}
			1 & 0 & 0 & 0\\
			0 & 1 & 0 & 0\\
			0 & 0 & \cos\delta_2 & \sin\delta_2 \\
			0 & 0 & -\sin\delta_2 &\cos\delta_2 \\
		\end{pmatrix} 
		\begin{pmatrix}
			1 & 0 & 0 & 0\\
			0 & \cos\delta_1 & 0 & -\sin\delta_1 \\
			0 & 0 & 1 & 0 \\
			0 & \sin\delta_1 & 0 &\cos\delta_1 \\
		\end{pmatrix} 
		\begin{pmatrix}
			1\\
			1 \\
			0 \\
			0 \\
		\end{pmatrix} \\&=
		\begin{pmatrix}
			1 & 0 & 0 & 0\\
			0 & 1 & 0 & 0\\
			0 & 0 & \cos\delta_2 & \sin\delta_2 \\
			0 & 0 & -\sin\delta_2 &\cos\delta_2 \\
		\end{pmatrix} 
		\begin{pmatrix}
			1\\
			\cos\delta_1 \\
			0 \\
			\sin\delta_1  \\
		\end{pmatrix} \\&=
		\begin{pmatrix}
			1\\
			\cos\delta_1 \\
			\sin\delta_1\sin\delta_2 \\
			\sin\delta_1 \cos\delta_2 \\
		\end{pmatrix} \\&\overset{\delta_2=+\frac{\pi}{2}}{=}
		\begin{pmatrix}
			1\\
			\cos\delta_1  \\
			\sin\delta_1 \\
			0 \\
		\end{pmatrix}  
	\end{align}
	
	with $\delta_1$ the retardance of LCVR\,1 and $\delta_2$ the retardance of LCVR\,2. When LCVR\,2 is set to a retardance of $\delta_2=+\pi/2$ for 3D-PLI measurements, the generated polarization is linear and its polarization angle spins counter-clockwise with increasing $\delta_1$ around the plane of the Poincaré sphere.
	
	A non-ideal PSG can deviate from the ideal values for $\delta_1$ and $\delta_2$, usually because of a minor voltage offset for LCVR\,1 and LCVR\,2. While a deviation from $\delta_1$ only results in a global phase shift, a deviation for $\delta_2$ causes elliptical polarization that propagates through the whole system. When $(\delta_2-\pi/2)=h_2\approx0$, a Taylor approximation to the 1st order with
	\begin{align}
		\cos(\delta_2\approx \frac{\pi}{2}) &= - (\delta_2-\frac{\pi}{2}) + ... \approx - h_2\\
		\sin(\delta_2\approx \frac{\pi}{2}) &= 1 - ... \approx 1
	\end{align}
	
	describes the Stokes vector $\vec{S}_{\textrm{real}}$ as a sum of the ideal vector $\vec{S}_{\textrm{PSG}}$ and a deviation $\vec{S}_{h2}$:
	\begin{align}
		\label{appendix:eq:stokes-taylor}
		\vec{S}_{\textrm{real}} &= \begin{pmatrix}
			1\\
			\cos\delta_1 \\
			\sin\delta_1\sin\delta_2 \\
			\sin\delta_1 \cos\delta_2 \\
		\end{pmatrix} \approx\begin{pmatrix}
			1\\
			\cos\delta_1 \\
			\sin\delta_1\\
			0 \\
		\end{pmatrix} -\begin{pmatrix}
			0\\
			0 \\
			0\\
			h_2\sin\delta_1\\
		\end{pmatrix}=\vec{S}_{\textrm{PSG}} + \vec{S}_{h2}
	\end{align}
	
	With the matrices for an ideal retarder and horizontal linear polarization filter at azimuthal angles $\theta$, the ideal PSA can be described by~\cite{Goldstein2011}:
	\begin{align}
		&M_{\textrm{PSA}}\\&=
		M_{\textrm{LP}}(\theta=0^\circ)\cdot M(\delta_4, \theta=45^\circ)\cdot M(\delta_3, \theta=0^\circ)
		\\=&	\frac{1}{2}\begin{pmatrix}
			1 & 1 & 0 & 0 \\
			1 & 1 & 0 & 0 \\
			0 & 0 & 0 & 0 \\
			0 & 0 & 0 & 0 \\
		\end{pmatrix}\begin{pmatrix}
			1 & 0 & 0 & 0\\
			0 & \cos\delta_4 & 0 & -\sin\delta_4 \\
			0 & 0 & 1 & 0 \\
			0 & \sin\delta_4 & 0 &\cos\delta_4 \\
		\end{pmatrix} 	\begin{pmatrix}
			1 & 0 & 0 & 0\\
			0 & 1 & 0 & 0\\
			0 & 0 & \cos\delta_3 & \sin\delta_3 \\
			0 & 0 & -\sin\delta_3 &\cos\delta_3 \\
		\end{pmatrix} 
		\\=&		\frac{1}{2}	\begin{pmatrix}
			1 & \cos\delta_4 & \sin\delta_4\sin\delta_3 & -\sin\delta_4\cos\delta_3\\
			1 &	\cos\delta_4 & \sin\delta_4\sin\delta_3 & -\sin\delta_4\cos\delta_3\\
			0 & 0 & 0 & 0\\
			0 & 0 & 0 & 0\\
		\end{pmatrix}\\\overset{\delta_3=0}{ \overset{\delta_4=\pm\frac{\pi}{2}}{=}}&
		\frac{1}{2}	\begin{pmatrix}
			1 & 0 & 0 & \mp1\\
			1 &	0 & 0 & \mp1\\
			0 & 0 & 0 & 0\\
			0 & 0 & 0 & 0\\
		\end{pmatrix}
	\end{align}
	
	with $\delta_3=0$ and $\delta_4=\pm\pi/2$ when analyzing either right- or left-circular polarization in 3D-PLI.
	
	A non-ideal PSA can deviate from the ideal values for $\delta_3$ and $\delta_4$, usually because of a minor voltage offset for LCVR\,3 and LCVR\,4, respectively. When $(\delta_4-\pi/2)=h_4$ and $(\delta_3-\pi/2)=h_3$, a Taylor approximation to the 1st order with:
	\begin{align}
		\cos(\delta_4\approx\pm\frac{\pi}{2}) &= \mp (\delta_4-\frac{\pi}{2}) + ... \approx \mp h_4\\
		\sin(\delta_4\approx\pm\frac{\pi}{2}) &= \pm 1 - ... \approx \pm1\\
		\cos(\delta_3\approx0) &= 1- ... \approx 1\\
		\sin(\delta_3\approx0) &= (\delta_3-0) - ...\approx h_3
	\end{align}
	
	describes the whole \textcolor{black}{Mueller} matrix for the PSA $M_{\textrm{real}}$ as a sum of two matrices, one for the ideal matrix $M_{\textrm{PSA}}$ and a deviation $M_{h3, h4}$:
	\begin{align}
		\label{appendix:eq:mueller-taylor}
		&M_{\textrm{real}}\\&=
		M_{\textrm{LP}}(\theta=0^\circ)\cdot M(\delta_4, \theta=45^\circ)\cdot M(\delta_3, \theta=0^\circ)
		\\=&		\frac{1}{2}	\begin{pmatrix}
			1 & \cos\delta_4 & \sin\delta_4\sin\delta_3 & -\sin\delta_4\cos\delta_3\\
			1 &	\cos\delta_4 & \sin\delta_4\sin\delta_3 & -\sin\delta_4\cos\delta_3\\
			0 & 0 & 0 & 0\\
			0 & 0 & 0 & 0\\
		\end{pmatrix}\\\approx&
		\frac{1}{2}	\begin{pmatrix}
			1 & 0 & 0& \mp1\\
			1 &	0& 0 & \mp1\\
			0 & 0 & 0 & 0\\
			0 & 0 & 0 & 0\\
		\end{pmatrix}+
		\frac{1}{2}	\begin{pmatrix}
			0 & \mp h_4 &\pm h_3 & 0\\
			0 &	\mp h_4 & \pm h_3 & 0\\
			0 & 0 & 0 & 0\\
			0 & 0 & 0 & 0\\
		\end{pmatrix}
		\\=&M_{\textrm{PSA}}+M_{h3, h4}\end{align}
	
	It is important to note that $h_4$ is not necessarily equal for $\delta_4\approx\pi/2$ and $\delta_4\approx-\pi/2$, i.e.\,can be different for right- and left-circular polarization because different voltage settings are applied to LCVR\,4. However, $h_3$ remains equal because the setting for LCVR\,3 $\delta_3\approx0$ is kept for both cases.
	
	\subsection*{Error propagation in 3D-PLI}
	
	In the following, the error propagation in 3D-PLI caused by non-ideal optical elements in a \textcolor{black}{Mueller} polarimeter is calculated, based on \textcolor{black}{Mueller} calculus and Taylor approximations of the first order. The mathematical results aid in identifying the sources of non-ideal behavior.
	
	\textcolor{black}{Here, we assume that an offset in the retardance of the LCVRs is the primary source of systematic errors. In comparison, the azimuthal angles of the LCVRs are of secondary importance. The main reason for assuming ideal angles lies in the hardware: azimuthal angles can be set with high precision using rotation mounts. In contrast, the retardance of the LCVRs must first be characterized to relate voltage settings to actual retardance values. Furthermore, the voltage–retardance curves exhibit inflection points at retardance values of $\pm\frac{\pi}{2}$ (i.e., when operated as quarter-wave plates), where the curves are particularly steep and thus highly sensitive to small errors. This increases the likelihood of deviations from the ideal retardance. By contrast, rotation mounts offer straightforward, linear, and precise adjustment of angular settings.}
	
	The \textcolor{black}{Mueller} matrix for brain tissue Eq.\,(\ref{appendix:eq:tissue-ret}), the ideal \textcolor{black}{Mueller} matrix of the PSA $M_{\textrm{PSA}}$ and the ideal Stokes vector of the PSG $\vec{S}_{\textrm{PSG}}$ can be applied to calculate the Stokes vector $\vec{S}_{\textrm{PLI}}$ for 3D-PLI which directly corresponds to the measured intensity. 
	
	In 3D-PLI, $\delta_1$ is variable, $\delta_2=\pi/2$, $\delta_3=0$ and $\delta_4=\pm\pi/2$: 
	\begin{align}
		&\vec{S}_{\textrm{PLI}}\\& = 	M_{\textrm{PSA}}\cdot M_{\textrm{tissue}}\cdot\vec{S}_{\textrm{PSG}}\\&	= 	M_{\textrm{PSA}}	
		\begin{pmatrix}
			1 & 0 & 0 & 0\\
			0 &	\cos^2(2\phi)+\sin^2(2\phi)\cos\delta & \cos(2\phi)\sin(2\phi)(1-\cos\delta) & -\sin(2\phi)\sin\delta\\
			0 & \cos(2\phi)\sin(2\phi)(1-\cos\delta) & \sin^2(2\phi)+\cos^2(2\phi)\cos\delta & +\cos(2\phi)\sin\delta\\
			0 & +\sin(2\phi)\sin\delta & -\cos(2\phi)\sin\delta &\cos\delta\\
		\end{pmatrix}
		\begin{pmatrix}
			1\\
			\cos(\delta_1)  \\
			\sin(\delta_1) \\
			0 \\
		\end{pmatrix} 		
		\\&= 	\frac{1}{2}	\begin{pmatrix}
			1 & 0 & 0 & \mp1\\
			1 &	0 & 0 & \mp1\\
			0 & 0 & 0 & 0\\
			0 & 0 & 0 & 0\\
		\end{pmatrix}
		\begin{pmatrix}
			1\\
			\cos\delta_1\cos^2(2\phi)+\sin^2(2\phi)\cos\delta + \cos(2\phi)\sin(2\phi)i(1-\cos\delta)\sin\delta_1 \\
			\cos(2\phi)\sin(2\phi)(1-\cos\delta)\cos\delta_1 + \sin^2(2\phi)+\cos^2(2\phi)\cos\delta\sin\delta_1 \\
			+\sin(2\phi)\sin\delta\cos\delta_1 -\cos(2\phi)\sin\delta\sin\delta_1 \\
		\end{pmatrix} 
		\\&= 
		\frac{1}{2}
		\begin{pmatrix}
			1 \pm \sin\delta\sin(\delta_1-2\phi) \\
			1 \pm \sin\delta\sin(\delta_1-2\phi)\\
			0\\
			0\\
		\end{pmatrix}
	\end{align}
	
	The first element of the Stokes vector $\vec{S}_{\textrm{PLI}}$ is directly equivalent to the 3D-PLI signal but with $\delta_1\hat{=}2\rho$
	\begin{align}
		\label{appendix:eq:pli-intensity-ideal}
		I_{\textrm{PLI}}= \frac{1}{2}\left(1 \pm \sin\delta\sin(\delta_1-2\phi)\right)
	\end{align}
	
	With the Taylor approximation in Eq.\,(\ref{appendix:eq:stokes-taylor}) and the abbreviations $v_\phi=\sin(2\phi)$, $w_\phi=\cos(2\phi)$, the non-ideal Stokes vector $\vec{S}_{\textrm{PLI'}}$ for a non-ideal PSG and an ideal PSA can be calculated as:
	\begin{align}
		&\vec{S}_{\textrm{PLI'}}\\& = 	M_{\textrm{PSA}}\cdot M_{\textrm{tissue}} \cdot	(\vec{S}_{PSG} + \vec{S}_{h2})\\& =
		\vec{S}_{\textrm{PLI}} + M_{\textrm{PSA}}M_{\textrm{tissue}}	\vec{S}_{h_2}\\& =\vec{S}_{\textrm{PLI}} -	\frac{1}{2}	\begin{pmatrix}
			1 & 0 & 0 & \mp1\\
			1 &	0 & 0 & \mp1\\
			0 & 0 & 0 & 0\\
			0 & 0 & 0 & 0\\
		\end{pmatrix}\begin{pmatrix}
			1 & 0 & 0 & 0\\
			0 &	w^2_\phi+v^2_\phi\cos\delta & w_\phi v_\phi(1-\cos\delta) & -v_\phi\sin\delta\\
			0 & w_\phi v_\phi(1-\cos\delta) & v^2_\phi+w^2_\phi\cos\delta & +w_\phi\sin\delta\\
			0 & v_\phi\sin\delta & -w_\phi\sin\delta &\cos\delta\\
		\end{pmatrix} \begin{pmatrix}
			0\\
			0 \\
			0\\
			h_2\sin(\delta_1)\\
		\end{pmatrix}\\&=	\frac{1}{2}	\left[
		\begin{pmatrix}
			1 \pm \sin\delta\sin(\delta_1-2\phi) \\
			1 \pm \sin\delta\sin(\delta_1-2\phi)\\
			0\\
			0\\
		\end{pmatrix}\pm	
		\begin{pmatrix}
			h_2\cos\delta\sin\delta_1\\
			h_2\cos\delta\sin\delta_1\\
			0\\
			0\\\end{pmatrix}\right]\\&= \vec{S}_{\textrm{PLI}}+\vec{S}_{\textrm{LCVR2}}(\delta, \delta_1)
	\end{align}
	
	The ideal 3D-PLI intensity curve in Eq.\,(\ref{appendix:eq:pli-intensity-ideal}) is now modulated by a function with an amplitude $h_2$ that depends on the fiber retardance $\delta$, the retardance $\delta_1$ of LCVR\,1 but not the fiber direction $\phi$:
	\begin{equation}
		I_{\textrm{PLI'}} = \frac{1}{2}\left(	1 \pm \sin\delta\sin(\delta_1-2\phi)\right)\pm 	\frac{h_2}{2}\cos\delta\sin(\delta_1).
	\end{equation}
	
	Using the Taylor approximation in Eq.\,(\ref{appendix:eq:mueller-taylor}) and the abbreviations $v_\phi=\sin(2\phi)$, $w_\phi=\cos(2\phi)$, the non-ideal Stokes vector $\vec{S}_{\textrm{PLI'}}$ for an ideal PSG and a non-ideal PSA can be calculated as:
	\begin{align}		
		&\vec{S}_{\textrm{PLI'}}\\& = (	M_{\textrm{PSA}}+M_{h3, h4})\cdot M_{\textrm{tissue}} \cdot 	\vec{S}_{\textrm{PSG}}\\& =\vec{S}_{\textrm{PLI}} + M_{h3, h4} \cdot M_{\textrm{tissue}}\cdot 	\vec{S}_{\textrm{PSG}}\\& =
		\vec{S}_{\textrm{PLI}}+\frac{1}{2}	\begin{pmatrix}
			0 & \mp h_4 &\pm h_3 & 0\\
			0 &	\mp h_4 & \pm h_3 & 0\\
			0 & 0 & 0 & 0\\
			0 & 0 & 0 & 0\\
		\end{pmatrix} \begin{pmatrix}
			1\\
			\cos\delta_1w^2_\phi+v^2_\phi\cos\delta + w_\phi v_\phi(1-\cos\delta)\sin\delta_1 \\
			w_\phi v_\phi(1-\cos\delta)\cos(\delta_1) + v^2_\phi+w^2_\phi\cos\delta\sin\delta_1 \\
			v_\phi\sin\delta\cos\delta_1 - w_\phi\sin\delta\sin\delta_1 \\
		\end{pmatrix} \\& =
		\vec{S}_{\textrm{PLI}}\mp\frac{h_4 }{2}
		\begin{pmatrix}
			\cos\delta_1w^2_\phi+v^2_\phi\cos\delta + w_\phi v_\phi(1-\cos\delta)\sin\delta_1 \\
			\cos\delta_1w^2_\phi+v^2_\phi\cos\delta + w_\phi v_\phi(1-\cos\delta)\sin\delta_1 \\
			0 \\
			0\\
		\end{pmatrix}\\\quad&\pm\frac{h_3}{2}
		\begin{pmatrix}
			w_\phi v_\phi(1-\cos\delta)\cos\delta_1 + v^2_\phi+w^2_\phi\cos\delta\sin\delta_1 \\
			w_\phi v_\phi(1-\cos\delta)\cos\delta_1 + v^2_\phi+w^2_\phi\cos\delta\sin\delta_1\\
			0 \\
			0\\
		\end{pmatrix} \\&=	\vec{S}_{\textrm{PLI}}+\vec{S}_{\textrm{LCVR3}}(\phi, \delta, \delta_1)+\vec{S}_{\textrm{LCVR4}}(\phi, \delta, \delta_1)
	\end{align}
	
	The ideal 3D-PLI intensity curve in Eq.\,(\ref{appendix:eq:pli-intensity-ideal}) is now modulated by two functions with amplitudes $h_3$ and $h_4$ that depend on the fiber retardance $\delta$, the retardance of LCVR\,1 $\delta_1$, and fiber direction $\phi$:
	\begin{align}
		I_{\textrm{PLI'}} &=\frac{1}{2}\left(1 \pm \sin\delta\sin(\delta_1-2\phi)\right) \\& \mp\frac{ h_4 }{2}\left(	\cos\delta_1w^2_\phi+v^2_\phi\cos\delta + w_\phi v_\phi(1-\cos\delta)\sin\delta_1\right) \\&\pm \frac{h_3}{2} \left(	w_\phi v_\phi(1-\cos\delta)\cos\delta_1 + v^2_\phi+w^2_\phi\cos\delta\sin\delta_1 \right)
	\end{align}
	
	With $v^2_\phi=\sin^2(2\phi)=\frac{1}{2}(1-\cos4\phi)$ and $w^2_\phi=\cos^2(2\phi)=\frac{1}{2}(1+\cos4\phi)$, a higher-order frequency is introduced into the signal that depends on the fiber direction.
	
	The transmitted Stokes vector $\vec{S}_{\textrm{PSA'}}$ for a non-ideal PSG and a non-ideal PSA can be calculated:
	\begin{align}
		\vec{S}_{\textrm{out}}& = (	M_{\textrm{PSA}}+M_{h3, h4})\cdot M_{\textrm{tissue}} \cdot (\vec{S}_{\textrm{PSG}}+\vec{S}_{h2})\\& =\vec{S}_{\textrm{PLI}} + \vec{S}_{\textrm{LCVR2}} +\vec{S}_{\textrm{LCVR3}}+ \vec{S}_{\textrm{LCVR4}} +M_{\textrm{h3, h4}}\cdot M_{\textrm{tissue}}\cdot \vec{S}_{\textrm{h2}}
	\end{align}
	
	which is the sum of previous results and a higher-order term. However, the multiplication of the very small factors $h_2$, $h_3$ and $h_4$ can be neglected.
	
	\begin{figure}[h]
		\centering
		\includegraphics[width=0.9\linewidth]{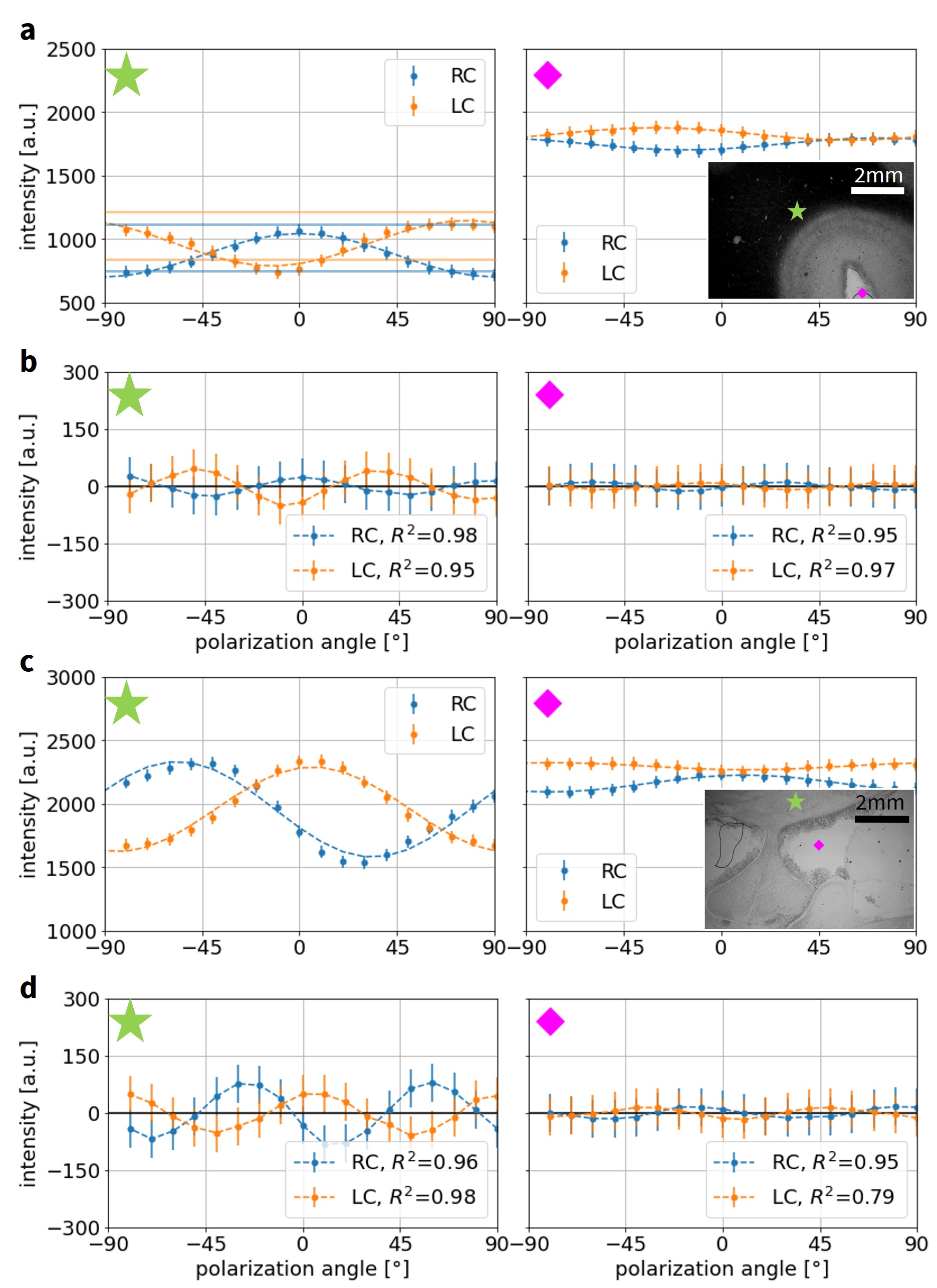}
		\caption{Fourier fit and residuals for exemplary pixels. 3D-PLI measurements with a left-handed circular (LC) and right-handed circular (RC) setting of the PSA are compared. (\textbf{a}) Fourier fit for the human brain section (low transmittance, high scattering). The two exemplary locations are marked in the transmittance map. (\textbf{b}) Corresponding residuals and $R^2$. (\textbf{c}) Fourier fit for the vervet monkey brain section (high transmittance, low scattering). The two exemplary locations are marked in the transmittance map. (\textbf{d}) Corresponding residuals and $R^2$ values.}
		\label{fig:3dpli-fit}
	\end{figure}


\end{document}